\documentclass{article}

\usepackage{PRIMEarxiv}

\usepackage[utf8]{inputenc}
\usepackage[T1]{fontenc}
\usepackage[dvipsnames]{xcolor}
\usepackage{amsmath,amssymb}
\usepackage{microtype,xspace,graphicx,fancyvrb,multirow}
\usepackage{booktabs,array,underscore,relsize,enumitem,tabularx}
\usepackage{subcaption}
\usepackage{tcolorbox,xstring,adjustbox}
\usepackage[sort]{cite}
\usepackage[breaklinks,colorlinks,citecolor=blue,urlcolor=black,linkcolor=blue]{hyperref}
\usepackage{iftex}
\ifPDFTeX
  \input{glyphtounicode}
\fi

\definecolor{lightgray}{rgb}{0.91,0.91,0.91}
\definecolor{lightblue}{rgb}{0.8627,0.9176,0.9686}

\graphicspath{{fig/}}

\newcommand{\sys}{\mbox{\textsc{SBind}}\xspace}
\newcommand{\sbridge}{\mbox{SBridge}\xspace}
\newcommand{\cc}[1]{\mbox{\smaller[0.5]\texttt{#1}}}
\def\Snospace~{\S{}}

\newcommand{\appautoref}[1]{\hyperref[#1]{Appendix~\ref{#1}}}
\newcommand{\PP}[1]{%
  \begingroup
    \edef\temp{\detokenize{#1}}%
    \IfEndWith{\temp}{.}{\noindent\textbf{#1}}{\noindent\textbf{#1.}}%
  \endgroup
}
\newcommand{\squishlist}{%
  \begin{itemize}[noitemsep,nolistsep,leftmargin=10pt]
  \setlength{\itemsep}{0pt}
}
\newcommand{\squishend}{\end{itemize}}
\newcommand{\summarybox}[1]{%
  \begin{tcolorbox}[colback=blue!3!white,boxrule=0.25mm,sharp corners,
    colframe=black,top=0pt,bottom=0pt,left=0pt,right=0pt]
    \textbf{Summary:}#1
  \end{tcolorbox}
}

\title{Behavioral Containment: Source-to-Binary Function Matching for 1-Day Vulnerability Detection}
\author{
  Yifei Yang\\
  Nanchang University and Nanyang Technological University \\
  \texttt{yifeiyang@ncu.edu.cn} \\
  \And
  Xuejiao Luo \\
  Nanyang Technological University and Beijing University of Posts and Telecommunications \\
  \texttt{luoxj@bupt.edu.cn} \\
   \And
  Penghui Li \\
  Nanyang Technological University \\
  \texttt{penghui.li@ntu.edu.sg} \\
}
\date{}

\hypersetup{
  pdftitle={Behavioral Containment: Source-to-Binary Function Matching for 1-Day Vulnerability Detection},
  pdfauthor={Anonymous Authors},
  pdfsubject={Source-to-binary function matching and patch presence detection}
}

\begin{document}
\maketitle

\begin{abstract}
Vulnerabilities in reused open-source code often propagate into deployed binaries.
Detecting them requires source-to-binary function matching, which locates the affected source function in a binary.
Applying prior binary-to-binary approaches to this task would require compiling the source code with the same build configuration as the target binary, which is usually unavailable.
Decompiling the binary does not make prior source-to-source approaches work either, because decompiled code differs substantially from the original source code.

We observe that compilation transforms the code of a function but largely preserves its behavior.
A binary function that implements a source function therefore contains its behavior, a property we call \emph{behavioral containment}.
Building on this, we present \sys, which decompiles the binary and checks behavioral containment instead of comparing code.
\sys represents each function by behavioral paths, sequences of normalized operations from its entry to its exits, and aligns source paths with binary paths in order.
To detect 1-day vulnerabilities, \sys compares the binary with the vulnerable and the patched source only on the paths through the changed lines.
On stripped binaries built with two compilers, two optimization levels, and four architectures, \sys ranks the correct function first in 77.87\% of 32,008 searches, against 45.69\% for the state-of-the-art approach \sbridge.
On 26 1-day CVEs, \sys reaches an F1 score of 0.70 on stripped binaries, against 0.51 for \sbridge.
Our replication of \sbridge also revealed several issues in its implementation and evaluation, and its authors have acknowledged them.

\end{abstract}


\section{Introduction}\label{sec:intro}

Open-source software is widely incorporated into commercial products, firmware, and other deployed systems~\cite{cisa2024oss}.
As a result, vulnerabilities in reused open-source components can propagate into downstream system binaries~\cite{cisa2024oss}.
To mitigate such supply-chain risks, security analysts scan deployed binaries to determine whether they contain vulnerable code and whether known security patches are present~\cite{duan2017oss,zhang2018precise}.

Code similarity is a foundational technique for securing software supply chains~\cite{woo2022movery,he2024sog,yang2026sbridge}.
Existing approaches mainly perform binary-to-binary or source-to-source code similarity analysis.
Source-to-source matching requires the target to be available as source code, which is not always the case for deployed software.
Binary-to-binary matching remains a common solution when reference binaries (e.g., ones containing known vulnerable code) are available~\cite{li2023libam,tang2020libdx,xu2023pem,yang2022modx,yang2026sbridge}.
However, the reference binaries must match the build configuration of the target, and constructing and maintaining them for every configuration can incur substantial practical costs~\cite{yang2026sbridge}.

We instead focus on \emph{source-to-binary} matching~\cite{duan2017oss,yuan2019b2sfinder,jiang2024binaryai,yang2026sbridge}, which determines whether a target binary is similar to the given source code.
Source-level vulnerability evidence is widely available through CVE reports, vulnerability databases, security advisories, and upstream patches~\cite{woo2021v0finder,woo2022movery,woo2023v1scan}.
Source-to-binary matching lets analysts use this evidence directly, without a reference binary that matches the unknown build configuration of the target.
A 1-day vulnerability is publicly disclosed and fixed, yet it may remain in deployed binaries~\cite{duan2017oss,woo2023v1scan}.
Our goal is to determine whether a stripped binary is vulnerable or patched, given the patch and the source code of the vulnerable function.
Source-to-binary function matching serves this goal by first locating this function in the binary.

Source-to-binary matching is difficult because source and binary code expose program behavior through fundamentally different representations.
First, compilation can remove or rewrite names, types, control flow, and operations, making direct source-to-binary comparison unreliable~\cite{aho2006compilers}.
Second, binaries built for different architectures, compiler settings, or optimization levels can differ substantially even when they are built from the same source code~\cite{aho2006compilers,he2024sog}.
Third, function inlining can place code from a source function inside the binary function of a caller, which we call its host function, so this code may occupy only part of a binary function~\cite{yang2026sbridge}.

Prior source-to-binary approaches remain limited in addressing these challenges.
B2SFinder~\cite{yuan2019b2sfinder} identifies coarse-grained reused software components (e.g., third-party libraries) but does not perform fine-grained function matching, which is important for locating vulnerable code and verifying patches.
Its features are also sensitive to compiler-induced variation.
BinaryAI~\cite{jiang2024binaryai} improves upon B2SFinder with function-level embeddings, but function inlining still causes errors.
\sbridge reports outperforming BinaryAI on its benchmark~\cite{yang2026sbridge}.
It decomposes source and binary functions into control blocks, such as condition and call blocks, and uses block-level matches as the basis for measuring function similarity~\cite{yang2026sbridge}.
However, it ignores the order of these blocks, so candidates that share blocks with the source function are difficult to distinguish.
Furthermore, as we show in our replication study, its implementation and evaluation contain several issues that affect the reported results (\autoref{sec:eval-rq1}).
For example, for one source function, \sbridge ranked 185 of the 534 candidate binary functions tied in first place, and its evaluation counted the match as correct because the correct function was among them.

Our key insight is that a binary function corresponding to a source function should exhibit the \emph{same essential behavior}, even when their local representations differ.
Since function inlining may embed the source behavior in a host function together with other operations, the matching requirement can be relaxed to \emph{behavioral containment}.
For example, a function that releases an XML entity checks the entity, releases its child nodes, releases each of its strings unless the document shares the string, and finally releases the entity.
The compiler may implement these steps differently, but the binary still performs them.

Realizing this intuition is non-trivial for two reasons.
First, a function may exhibit multiple behaviors across different inputs and control-flow outcomes.
It is difficult to define which behaviors are essential and represent them consistently across source and binary code~\cite{aho2006compilers,duan2017oss,yuan2019b2sfinder,jiang2024binaryai}.
Second, the resulting representations must be compared and matched across compiler transformations~\cite{yuan2019b2sfinder,jiang2024binaryai,he2024sog,yang2026sbridge}.
This comparison must also tolerate differences in code granularity and inlined code inside host functions~\cite{yang2026sbridge}.

We present \sys, a source-to-binary matching system that addresses these challenges through behavioral path alignment.
To address the representation challenge, \sys defines a behavioral path as an ordered sequence of semantically meaningful operations from a function entry to one of its exits.
An exit marks a possible function outcome such as a return or termination, and the operations along the path to it jointly determine that outcome.
Because compilation largely preserves these operations and their order, \sys represents a function by multiple behavioral paths, which capture its essential behavior across different outcomes.
To address the matching challenge, \sys estimates behavioral containment by aligning source and binary paths while tolerating additional binary operations.
The resulting path correspondences are aggregated into a function-level similarity score in which paths that cover more behavior carry greater weight.
\sys then refines the resulting scores with the calling context of each function.

Over all 32,008 stripped queries, with ties counted as misses, \sys ranks the correct function first for 77.87\%, against 45.69\% for \sbridge, and leads for every compiler, optimization level, and architecture.
It still leads when \sbridge is measured as in its original evaluation, with 79.90\% against 71.23\%.
In 1-day detection, the F1 score of \sys on stripped binaries is 0.70, against 0.51 for \sbridge.

Our contributions are summarized as follows.
\squishlist
\item We formulate source-to-binary function matching as \emph{behavioral containment}, which also covers inlined code, and apply it to 1-day vulnerability detection in stripped binaries.
\item We present \sys, which defines behavioral paths, estimates behavioral containment by aligning source and binary paths, and refines the scores with the calling context of each function.
\item We conduct a systematic replication and audit of \sbridge, identifying implementation, ranking, and ground-truth issues that affect its reported results.
\item We evaluate source-to-binary function matching and downstream 1-day vulnerability detection on separate datasets.
\squishend

\section{Background}\label{sec:background}

\subsection{Compilation and Decompilation}\label{sec:background-compilation}

A compiler translates the source code of a program into machine code, and the resulting binary is what vendors ship and devices run~\cite{aho2006compilers}.
Compilation removes variable names and types, and it turns structure fields into constant memory offsets.
Optimizations restructure control flow, and remove and reorder operations, and their effects differ across compilers and optimization levels~\cite{ren2021unleashing}.
The same source compiled for different architectures, such as x86 and ARM, also yields different instructions, registers, and calling conventions~\cite{he2024sog}.
Deployed binaries are commonly stripped of their static symbol table, which removes the names of internal functions~\cite{yang2026sbridge}, and an analysis must then recover function boundaries from the machine code~\cite{andriesse2016indepth}.
We use \emph{source function} for a function as written in the source code and \emph{binary function} for a function recovered from a binary.
The inlining optimization copies the body of a called function into its caller, so one binary function can contain code of several source functions~\cite{jia2023inlining,yang2026sbridge}.
We call such a binary function the \emph{host function} of the inlined code.

Decompilation partially reverses this process by lifting the machine code of each recovered function into C-like code with variables and structured control flow~\cite{brumley2013phoenix,ghidra}.
We call this output \emph{decompiled code}. It resembles C but need not compile.
A decompiler that supports several architectures lifts their binaries into the same C-like form, so an analysis of decompiled code does not need to handle each instruction set separately.
The recovery is approximate.
Without symbols, the decompiler assigns synthetic names to functions and variables (e.g., \cc{FUN\_0013a1f0} and \cc{param\_1}), and the types and call arguments that it infers can differ from those in the source.
Because decompiled code is C-like, the same front end can parse source code and decompiled code into a code property graph (CPG), which combines the abstract syntax tree, the control flow graph, and data dependence edges~\cite{yamaguchi2014modeling}.

\subsection{Code Similarity}\label{sec:background-similarity}

Code similarity analysis determines whether two pieces of code, typically functions, implement the same functionality~\cite{he2024sog,yang2026sbridge}.
Each piece can be source or binary code, which gives source-to-source, binary-to-binary, and source-to-binary analysis.
In software supply chains, similarity analysis supports software composition analysis, which identifies the open-source components and libraries reused in a product~\cite{duan2017oss,yuan2019b2sfinder,tang2020libdx,li2023libam,jiang2024binaryai}.
We call a function that the patch of a vulnerability modifies a \emph{vulnerable function}.
Similarity analysis detects 1-day vulnerabilities by finding reused copies of the affected code and determining whether each copy is vulnerable or patched~\cite{woo2022movery,woo2023v1scan}.
Source-to-binary analysis lets an analyst use the source code and patches of disclosed vulnerabilities to examine deployed binaries whose source code is unavailable~\cite{yuan2019b2sfinder,jiang2024binaryai,yang2026sbridge}.

\section{Problem and Key Idea}
\label{sec:problem}

\subsection{Problem Formulation}\label{sec:problem-task}

\PP{Function matching}
In \emph{source-to-binary function matching}, an analyst has the C source code of a program and one or more stripped target binaries built from it.
The compiler and optimization level of the binaries are unknown.
Given a source function $s$ and a target binary $B$, a \emph{counterpart} of $s$ is a function of $B$ that contains the compiled code of $s$.
It is either a separate function or a host function with an inlined copy of $s$.
A \emph{source-to-binary query} (\emph{query} for short) searches $B$ for the counterparts of $s$.
We call the functions recovered from $B$ the \emph{candidates}.
The task is to rank the candidates by a matching score so that the counterparts of $s$ come first.
Through inlining, a query can have several counterparts.
It can also have no counterpart among the candidates, for example if function recovery misses it.

\PP{1-day vulnerability detection}
For a 1-day vulnerability, the analyst also has the vulnerable and the patched source code of its vulnerable function.
The target binary is built from the vulnerable or the patched source code of the program, but the analyst does not know which one.
Detection first runs a query to locate its counterpart in the binary.
It then compares this counterpart with the vulnerable and the patched source code to decide whether the binary is vulnerable or patched.

\subsection{Motivating Example}\label{s:problem-example}

\begin{figure}[t]
\centering
\definecolor{figa}{HTML}{1F5FBF}\definecolor{figb}{HTML}{A0208C}\definecolor{figc}{HTML}{9A5B00}\definecolor{figd}{HTML}{1A7F37}
\newcommand{\figA}[1]{\textcolor{figa}{#1}}\newcommand{\figB}[1]{\textcolor{figb}{#1}}\newcommand{\figC}[1]{\textcolor{figc}{#1}}\newcommand{\figD}[1]{\textcolor{figd}{#1}}
\definecolor{addbg}{HTML}{EEFBF1}\definecolor{delbg}{HTML}{FFF2F1}\definecolor{addfg}{HTML}{1A7F37}\definecolor{delfg}{HTML}{CF222E}
\newcommand{\codeskip}{9.5pt}%
\newcommand{\diffbg}[3]{\makebox[0pt][l]{\hspace*{-2pt}\textcolor{#1}{\rule[-3pt]{\dimexpr\linewidth-2pt\relax}{\codeskip}}}\textcolor{#2}{#3}}
\newcommand{\AddBG}{\diffbg{addbg}{addfg}{+}}\newcommand{\DelBG}{\diffbg{delbg}{delfg}{-}}
\renewcommand{\theFancyVerbLine}{\textcolor{gray}{\sffamily\fontsize{5}{6}\selectfont\arabic{FancyVerbLine}}}
\fvset{fontfamily=fvm,fontsize=\fontsize{6}{\codeskip}\selectfont,numbers=left,numbersep=4pt,xleftmargin=9pt,frame=single,framerule=0.4pt,framesep=1.5pt,commandchars=\@\^\~}
\begin{subfigure}[b]{0.505\textwidth}
\begin{Verbatim}
static void @textbf^xmlFreeEntity~(xmlEntityPtr entity) {
  xmlDictPtr dict = NULL;
  @figA^if (entity == NULL)~ return;
  if (entity->doc != NULL)
    @figA^dict = entity->doc->dict;~
  if ((entity->children) && (entity->owner == 1) &&
      (entity == (xmlEntityPtr) entity->children->parent))
    @figB^xmlFreeNodeList(entity->children);~
@DelBG^~ if (dict != NULL) {
@DelBG^~   if ((entity->name != NULL) &&
@DelBG^~       (!xmlDictOwns(dict, entity->name)))
@DelBG^~     xmlFree((char *) entity->name);
@DelBG^~   if ((entity->ExternalID != NULL) &&
@DelBG^~       (!@figC^xmlDictOwns(dict, entity->ExternalID)~))
@DelBG^~     @figC^xmlFree((char *) entity->ExternalID);~
@DelBG^~   ...
@DelBG^~ } else {
@DelBG^~   ...
@DelBG^~ }

@AddBG^~ if ((entity->name != NULL) &&
@AddBG^~     ((dict == NULL) || (!xmlDictOwns(dict, entity->name))))
@AddBG^~   xmlFree((char *) entity->name);
@AddBG^~ if (entity->ExternalID != NULL)
@AddBG^~   xmlFree((char *) entity->ExternalID);
@AddBG^~ ...
  @figC^xmlFree(entity);~
}
\end{Verbatim}
\caption{Source code of LibXML2 with the fix.}
\label{fig:motivation-src}
\end{subfigure}%
\hfill%
\begin{subfigure}[b]{0.485\textwidth}
\begin{Verbatim}
void @textbf^FUN_0013a1f0~(long param_1) {
  ...
  @figA^if (param_1 == 0)~ { return; }
  lVar4 = *(long *)(param_1 + 0x18);
  if (*(long *)(param_1 + 0x40) == 0) {
    ...
  } else {
    @figA^lVar1 = *(long *)(*(long *)(param_1 + 0x40) + 0x98);~
    if (((lVar4 != 0) && (*(int *)(param_1 + 0x80) == 1)) &&
        (*(long *)(lVar4 + 0x28) == param_1)) {
      @figB^xmlFreeNodeList();~
    }
    lVar4 = *(long *)(param_1 + 0x10);
    if (lVar1 != 0) {
      ppcVar3 = (code **)PTR_xmlFree_00265fa8;
      ...
      if ((*(long *)(param_1 + 0x60) != 0) &&
          (iVar2 = @figC^xmlDictOwns(lVar1)~, iVar2 == 0)) {
        @figC^(**ppcVar3)(*(undefined8 *)(param_1 + 0x60));~
      }
      ...
      goto LAB_0013a2d2;
    }
  }
  ...
LAB_0013a2d2:
  @figC^(**ppcVar3)(param_1);~
  return;
}
\end{Verbatim}
\caption{Decompiled code from a stripped GCC \cc{-O2} build of the vulnerable release.}
\label{fig:motivation-bin}
\end{subfigure}
\caption{Simplified \cc{xmlFreeEntity} function in LibXML2. Red and green backgrounds mark the lines that the fix removes and adds. Text of the same color corresponds across the two listings.}
\label{fig:motivation}
\end{figure}

\autoref{fig:motivation-src} shows the vulnerable function \cc{xmlFreeEntity} in LibXML2, with the patch highlighted.
The function releases an entity, a named piece of content that an XML document can reference, together with its strings.
In the vulnerable code, it checks the entity (line~3), reads the dictionary of the document (lines~4--5), and releases the child nodes of the entity (lines~6--8).
The dictionary is a string table shared across the document, and the vulnerable code stores the identifiers and short contents of an entity in it.
The function therefore releases each of the six strings of the entity only if the dictionary does not own it (lines~9--19 show two of them), and finally releases the entity (line~27).

Storing these strings in the dictionary causes the vulnerability.
To break a reference cycle between entities, the parser overwrites the first byte of the content of an entity.
If the content is a dictionary entry, this write corrupts the dictionary and can cause double frees~\cite{libxml2fix40304}.
The fix keeps only the name of an entity in the dictionary and gives the entity private copies of its other strings, including its content.
\cc{xmlFreeEntity} does not perform the write, but it releases these strings, so the patched version tests ownership only for the name (lines~21--26).
Because the faulty write is unchanged, only the code that the fix changes, such as \cc{xmlFreeEntity}, shows whether a binary is patched.

\PP{Compilation effects}
Compilation hides the function and rewrites many of its local features, such as names, expressions, and calls.
\autoref{fig:motivation-bin} shows the decompiled function, one of about 2,300 functions recovered from the binary.
Only callees that LibXML2 exports, such as \cc{xmlFreeNodeList} and \cc{xmlDictOwns}, keep their names.
The compiler lowers field accesses to pointer arithmetic with constant offsets (line~4).
Optimization also changes the control flow.
It splits the code on whether the entity has a document (lines~5--7), so the test of the child nodes appears in both branches (lines~9--10 and the code omitted in line~6).
It also keeps the address of the global function pointer \cc{xmlFree} in a local variable (line~15), so most calls to \cc{xmlFree} no longer show its name (lines~19 and~27).
Decompilation does not undo these changes and adds its own.
The calls to \cc{xmlFreeNodeList} and \cc{xmlDictOwns} show fewer arguments than the source (lines~11 and~18), and the ownership test becomes a comma expression (line~18).

\PP{Vulnerable or patched}
Locating the function does not reveal whether the binary is patched.
The vulnerable and the patched function share most of their code.
They differ only in the highlighted lines of \autoref{fig:motivation-src}, where the fix removes five of the six ownership tests.
A binary function that resembles one version therefore also resembles the other.
Detection must compare the code of these lines, which compilation rewrites as well.

\subsection{Limitations of Prior Work}\label{sec:problem-prior}
Source code and binary code are two different representations of a program, so similarity checking must first bring them into a common representation~\cite{yuan2019b2sfinder,jiang2024binaryai,yang2026sbridge}.
Decompiling the binary into C-like code brings the problem close to source-to-source similarity.
Decompilation needs no build configuration and lifts binaries of different architectures into the same C-like form.
However, existing source-to-source techniques cannot be reused directly, because decompiled code differs substantially from the source code and normalizing these differences away is difficult.
For example, the clone detector VUDDY~\cite{kim2017vuddy}, which compares normalized function bodies, fails to match a single query in our evaluation.

\PP{Source-to-binary approaches}
Prior source-to-binary approaches compare code through local features such as strings, constants, and code blocks~\cite{duan2017oss,yuan2019b2sfinder,yang2026sbridge}.
The state-of-the-art approach \sbridge~\cite{yang2026sbridge} decompiles the binary and also compares local features.
It splits the source function and the decompiled function into \emph{control blocks}, small units of code of one kind, such as a condition or a call.
It then matches the blocks of the two functions pairwise.
Its score grows with the share of source blocks that find a match and decreases with the number of binary blocks.
\sbridge thus treats a function as a set of local blocks and ignores their order.

In this example, few source blocks match the decompiled function.
Call blocks fail because the calls lose arguments in the binary (lines~11 and~18 of \autoref{fig:motivation-bin}), and \sbridge requires the number of arguments to agree.
Condition blocks fail because offsets replace the field accesses and comma expressions replace the ownership tests.
Short functions that release other structures, such as URI structures, match more blocks.
They have fewer binary blocks and thus a smaller penalty, so \sbridge ranks one of them first.
Yet the binary largely keeps the order of the source code.

\subsection{Our Key Idea}\label{sec:problem-idea}

Like \sbridge, we decompile the binary.
Our key insight is that a binary function that implements a source function still exhibits the behavior of the source function across compilers, optimization levels, and architectures.
By behavior, we mean the 
operations a function performs and their order.
Examples are condition tests, memory reads, and calls.
Because inlining can place this behavior in a host function, the binary function need not equal the source function but only contain its behavior.
We call this requirement \emph{behavioral containment}.

\autoref{fig:motivation} illustrates behavioral containment.
On the branch where the document has a dictionary, both functions perform the same operations in the same order.
The binary only adds operations, such as the load in line~4 and the copy of the function pointer in line~15 of \autoref{fig:motivation-bin}.
Its ownership test of the external identifier (lines~17--19) exists only in the vulnerable function, so behavior also shows that this binary is vulnerable.

\subsection{Technical Challenges}\label{sec:problem-challenges}

Realizing behavioral containment raises two challenges.

\PP{C1: Representing behavior}
A function behaves differently on different executions, depending on its inputs.
In \autoref{fig:motivation-src}, for example, the function returns immediately for a null entity and otherwise releases the entity and its strings.
Isolated blocks do not capture such behaviors, and a single representation of the whole function mixes them.
Enumerating all control-flow paths is infeasible, because their number grows exponentially with the number of branches.
The representation must therefore select a manageable set of behaviors that still characterizes the function.
It must also make the behaviors of source and decompiled code comparable.
Yet normalizing too much makes unrelated functions look alike.

\PP{C2: Checking containment}
Penalizing the additional code of a counterpart lowers the rank of a host function, whereas ignoring it favors large unrelated functions that contain many operations by chance.
The check must also separate similar candidates.
For 1-day detection, the check must finally separate the vulnerable and the patched function, which differ in only a few operations.

\section{Approach}\label{sec:approach}

\begin{figure}[t]
\centering
\includegraphics[width=\textwidth]{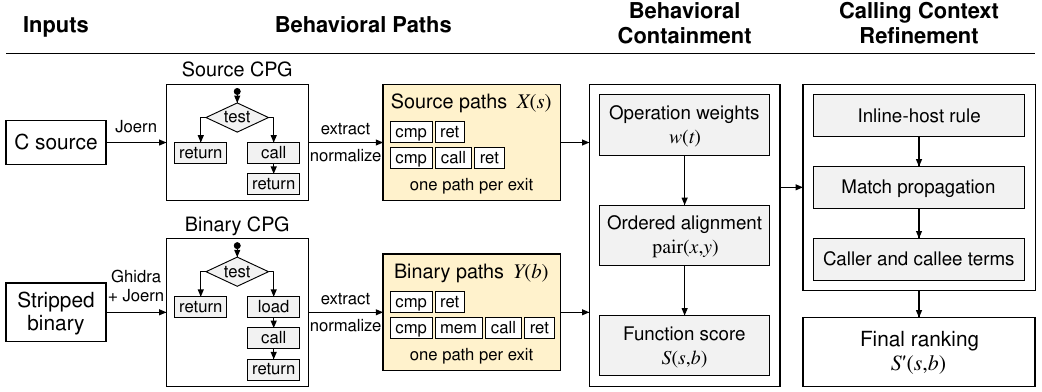}
\caption{Overview of function matching in \sys. 
}
\label{fig:approach-overview}
\end{figure}

For C1, \sys represents each function by normalized \emph{behavioral paths}, sequences of operations from the entry of the function to one of its exits.
For C2, \sys measures how much of each source path a binary path contains.
\autoref{fig:approach-overview} shows the function matching workflow of \sys.
The first stage decompiles the binaries and extracts the behavioral paths of each source and binary function (\autoref{sec:approach-repr}).
The second stage aligns source paths with binary paths to estimate behavioral containment (\autoref{sec:approach-scoring}).
The third stage refines the resulting scores with the calling context of each function (\autoref{sec:approach-consistency}).
For 1-day vulnerability detection, a further stage locates the vulnerable function in the binary and compares it with the vulnerable and the patched source code (\autoref{sec:approach-rq4}).

\subsection{Extracting Behavioral Paths}
\label{sec:approach-repr}

Each execution of a function follows one path through its control flow (C1).
\sys therefore represents a function by a set of control-flow paths~\cite{david2014tracelet,ding2019asm2vec}, each of which approximates one execution.
Paths suit the problem for three reasons.
First, a path keeps the operations that occur together and separates those of mutually exclusive branches. 
A set of CPG nodes, in contrast, mixes the operations of all branches.
Second, a path keeps the order of its operations.
Third, a patch changes operations on only some paths of a function, so paths separate the changed behavior from the shared behavior, which 1-day detection needs.

Among the possible paths,
\sys uses those from function entry to exit.
An exit marks one outcome of the function.
Entry-to-exit paths suit matching for three reasons.
First, a full path holds all operations that jointly determine one outcome, including the conditions that lead to it. 
Second, entries and exits are stable boundaries, because they can be identified independently in source and binary code.
Program points inside the function are not stable, because optimization splits and merges its basic blocks.
Third, a source path can match inlined code that forms the middle part of a longer binary path.

\subsubsection{Decompilation and Parsing}
We use Ghidra~\cite{ghidra} to recover the functions of a binary and lift each function into decompiled code.
The same C front end parses  source and decompiled code into CPGs with the same graph structure.
The source side uses preprocessed translation units, in which macros are expanded, and headers are included.
Variable, parameter, and field names are excluded from the representation, because compilation removes them.

\subsubsection{Path Extraction}\label{sec:approach-paths}
\sys extracts behavioral paths from the CPG of each function in three steps, which define the operations, identify the exits, and select one path per exit.
Each operation, such as a read, an assignment, a comparison, or a call, is one node of the CPG.
An exit is a return statement, a call to a non-returning function, or the end of a function without a return value.

\begin{figure}[t]
\centering
\begin{subfigure}[b]{\columnwidth}
\centering
\includegraphics[width=0.6\columnwidth]{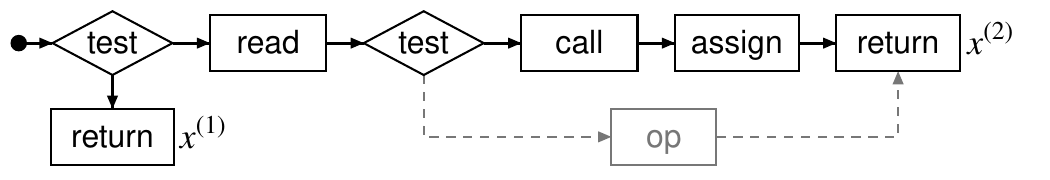}
\caption{Path extraction}
\label{fig:approach-alignment-a}
\end{subfigure}
\begin{subfigure}[b]{\columnwidth}
\centering
\includegraphics[width=0.6\columnwidth]{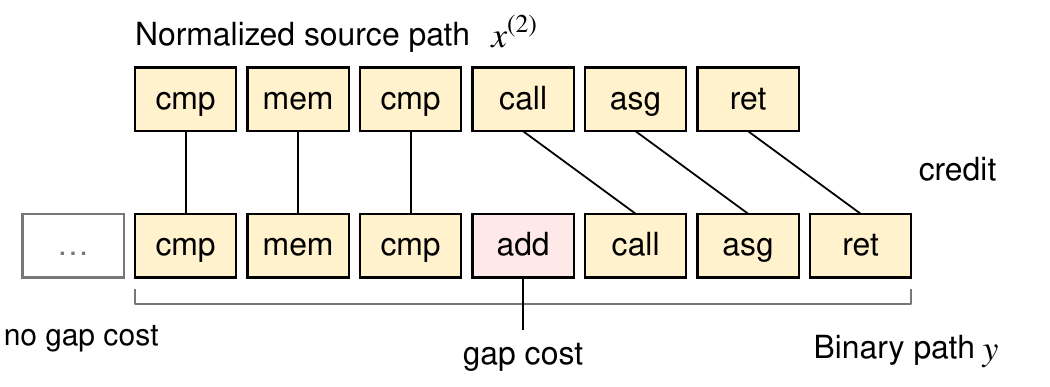}
\caption{Ordered path alignment}
\label{fig:approach-alignment-b}
\end{subfigure}
\caption{Path extraction and ordered alignment on a schematic example. 
In (a), the dashed path ends at the same exit as an extracted path and is not extracted.
}
\label{fig:approach-alignment}
\end{figure}

Along a path, each loop body is by default entered at most once, and a cycle of \cc{goto} statements is entered at most once at each of its entry points.
To avoid path explosion, \sys keeps one path per exit and thus one behavior per outcome of the function.
Among the paths to an exit, \sys keeps the longest by number of operations (\autoref{fig:approach-alignment-a}), which covers the most operations that lead to the outcome of this exit.
Choosing the longest path is a coverage heuristic, because length does not show that an operation matters.
A branch on no extracted path is not represented.
On the stripped binaries of the function matching dataset (\autoref{sec:eval-setup}), the extracted paths cover 85.4\% of all source operations and 76.0\% of all binary operations.
The patch presence test adds paths through the changed lines.
\sys extracts paths in the same way from source and decompiled code, and \autoref{sec:implementation} gives the limits of the extraction.
\appautoref{app:path} gives details.

\subsubsection{Operation Normalization}\label{sec:approach-normalization}
To make the paths of source and decompiled code comparable at the right level (C1), operation normalization maps each operation to zero or more \emph{normalized operations} from a fixed vocabulary of operators, calls, string literals, and integer constants, guided by two principles.
From here on, an operation means a normalized operation unless stated otherwise.
Normalization keeps what a stripped binary still contains and abstracts away only the differences that compilation and decompilation introduce systematically.

\autoref{tab:normalization} lists the main rules, each with an example from source code and from decompiled code.
Following the first principle, calls to library functions and to other functions in the dynamic symbol table keep their names, which a stripped binary retains for dynamic linking.
String literals and most integer constants are also kept.
Following the second principle, calls to internal functions map to one \emph{generic call} that does not identify the callee, as does a source call if no candidate of the binary calls a function of that name.
Reads of fields, array elements, and dereferenced pointers map to one memory access, because decompiled code often expresses all three as pointer arithmetic.
Compilers invert comparisons and the decompiler swaps their operands, so all equality comparisons map to one operation and all relational comparisons to another.
Finally, operations that carry little information or have no stable equivalent are dropped.
The constants 0 and 1, which appear on almost every path, are kept only where they are assigned to variables, returned, or passed as arguments, and constructs such as \cc{sizeof} and \cc{goto} are removed.

For the function in \autoref{fig:motivation}, the vulnerable source function has two paths, one for each exit, and its counterpart in the binary has three, because the compiler duplicates the release of the entity at an additional exit.
After normalization, the source paths have 3 and 68 operations, and the binary paths have 2, 77, and 120.

\begin{table}[t]
\centering
\caption{Main normalization rules.}
\label{tab:normalization}
\footnotesize
\setlength{\tabcolsep}{3pt}
\begin{tabularx}{0.5\columnwidth}{@{}Xlll@{}}
\toprule
Rule & Source & Decompiled & Normalization \\
\midrule
Load & \cc{e->doc} & \cc{*p} & \cc{mem} \\
Store & \cc{a[i] = x} & \cc{*p = x} & \cc{addr}, \cc{store} \\
Named call & \cc{f(x, y)} & \cc{f(x)} & \cc{call(f)} \\
Internal call & \cc{g(x)} & \cc{FUN\_0013a1f0()} & \cc{call(gen)} \\
Indirect call & \cc{(*fp)(x)} & \cc{(*pcVar1)(x)} & \cc{icall} \\
Fortified call & \cc{memcpy()} & \cc{\_\_memcpy\_chk()} & \cc{call(memcpy)} \\
Output call & \cc{printf()} & \cc{puts()} & \cc{call(printf)} \\
Equality & \cc{x != y}, \cc{!x} & \cc{x == y} & \cc{cmp} \\
Relation & \cc{x <= y} & \cc{y < x} & \cc{lt} \\
Compound & \cc{x += y}, \cc{x++} & \cc{x = x + y} & \cc{add}, \cc{asg} \\
Negative & \cc{\textasciitilde 8} & \cc{0xfffffff7} & \cc{int(-9)} \\
Subtraction & \cc{x - 4} & \cc{x + -4} & \cc{int(-4)}, \cc{add} \\
\bottomrule
\end{tabularx}
\end{table}

\subsection{Checking Behavioral Containment}\label{sec:approach-scoring}

With behaviors represented as paths, checking behavioral containment reduces to asking how much of a source path a binary path contains.
A symmetric similarity measure, such as the Jaccard index of two operation sets, does not fit this question, because it counts the additional code of a counterpart against it (C2).
\sys therefore measures containment, which tolerates additional binary operations, by the coverage of a source path in a binary path.
However, tolerating additional code favors large unrelated functions (C2).
\sys therefore follows three principles that separate a counterpart from such a function.

First, rare operations matter more.
A string literal or a call to a specific library function identifies a function much better than an assignment, so \sys weights each operation by its rarity (\autoref{sec:approach-idf}).
Second, order matters.
An unrelated function can perform the same operations as the source function in another order, whereas a counterpart keeps the order of the source function.
Set containment cannot distinguish the two because it ignores order.
\sys instead aligns the operations of two paths in order and penalizes matches that are scattered through a long binary path (\autoref{sec:approach-pair}).
Third, long paths matter more.
A long path carries more behavior than a short error path, and a short source path matches by chance more easily, so \sys weights each source path by its length when it combines the paths into a function score (\autoref{sec:approach-function}).

\subsubsection{Operation Weights}\label{sec:approach-idf}
\sys weights each operation $t$ by its inverse document frequency $w(t)$~\cite{sparckjones1972statistical} over the paths of all binaries under analysis (\appautoref{app:weights}).
Over the eight stripped target binaries of the vulnerability in \autoref{fig:motivation}, the return has weight 1.00 and the memory access has weight 1.24.
The calls to \cc{xmlDictOwns} and \cc{xmlFreeNodeList} have weights 5.58 and 7.07.

\subsubsection{Path-Pair Score}\label{sec:approach-pair}
Let $x=(x_1,\dots,x_n)$ and $y=(y_1,\dots,y_m)$ denote normalized source and binary paths, respectively, with weights $W(x)=\sum_i w(x_i)$ and $W(y)=\sum_j w(y_j)$.
An alignment $A$ is a set of index pairs $(i,j)$ that is strictly increasing in both indices (\autoref{fig:approach-alignment-b}).
The coverage of $x$ in $y$ is
\begin{equation}\label{eq:cov}
\begin{split}
  \mathrm{cov}(x,y) = \frac{1}{W(x)}\max_{A}\Bigl[&\sum_{(i,j)\in A}\mathrm{credit}(x_i,y_j)\\
  &-\lambda\sum_{j\in\mathrm{gap}(A)} w(y_j)\Bigr],
\end{split}
\end{equation}
where $\mathrm{gap}(A)$ contains the unmatched binary positions between the first and the last matched position.
Unmatched binary operations inside this range cost $\lambda=0.1$ times their weight, so a compact match scores higher than a match scattered through a large function.
The surrounding code of a host function lies outside this range and adds no gap cost.
The credit is $w(x_i)$ for identical operations.
If one operation is the generic call and the other a named call, the credit is $\kappa\,w(x_i)$ with $\kappa=0.4$.
This partial credit accounts for wrapper functions, such as the internal gnulib wrapper \cc{rpl\_free}, which the binary calls where the source calls \cc{free}.
No other pair of operations can match.
If $W(x)=0$, the coverage is 0.
Dynamic programming computes the best alignment in $O(nm)$ time~\cite{smith1981identification}.

The pair score multiplies the coverage by a length factor, because a long binary path matches source operations by chance more easily:
\begin{equation}\label{eq:pair}
  \mathrm{pair}(x,y) = \mathrm{cov}(x,y)\cdot\min\biggl(1,\frac{W(x)}{W(y)}\biggr)^{\beta}\,.
\end{equation}
A small exponent, $\beta=0.15$, limits this penalty.
In \autoref{fig:motivation}, the long source path selects the binary path with 120 operations.
The best alignment matches 49 of the 68 source operations and leaves 58 binary operations in its gap.
The coverage is 0.68, the length factor is 0.91, and the pair score is 0.62.

\subsubsection{Function Score}\label{sec:approach-function}
A source function $s$ has the path set $X(s)$, and a candidate $b$ has the path set $Y(b)$.
Each source path selects its best binary path, and several source paths may select the same binary path.
The function score weights each source path by its length $\ell(x)$, the number of operations on $x$ before normalization:
\begin{equation}\label{eq:func}
  S(s,b) = \frac{\sum_{x\in X(s)} \ell(x)\,\max_{y\in Y(b)}\mathrm{pair}(x,y)}{\sum_{x\in X(s)} \ell(x)}\,.
\end{equation}
\autoref{sec:approach-consistency} refines the function score into the matching score $S'(s,b)$.
In \autoref{fig:motivation}, the two source paths have 9 and 158 operations before normalization and pair scores of 0.64 and 0.62.
The counterpart scores 0.62 and ranks first among the 2,330 candidates, ahead of the next-best candidate at 0.45.
With equal operation weights, the counterpart stays first with 0.59, but the next-best candidate rises to 0.58.

\subsection{Refining with Calling Context}\label{sec:approach-propagation}\label{sec:approach-consistency}

The function score compares a source function with each candidate in isolation, which leaves two problems.
First, the length factor that guards against large unrelated functions also lowers the score of a host function that contains an inlined callee.
Second, similar candidates remain hard to separate (C2).
In \autoref{fig:motivation}, the counterpart ranks first only because LibXML2 exports its functions, such as \cc{xmlFreeNodeList} and \cc{xmlDictOwns}.
If calls to all exported functions of LibXML2 become generic calls, as if these functions were internal, the counterpart falls to rank 4.

\sys addresses both problems with the \emph{calling context} of a function, which consists of its callers and its callees~\cite{haler2004structual}.
Calls connect functions on both sides, so a confident match of one function also constrains its callers and callees.
The counterpart of a caller should call the matched binary function unless the compiler inlined it, and the counterpart of a callee should be called by it.
The third stage uses this evidence in three steps, based on the \emph{top candidate} of each source function, its highest-scoring candidate.

\subsubsection{Inline-Host Rule}
If a call on the paths of a source function $g$ targets a source function $f$, and $b$ is the top candidate of $g$, the compiler may have inlined $f$ into $b$.
In this case, \sys raises $S(f,b)$ to at least $\rho\, C(f,b)$, where $\rho=0.9$ and $C(f,b)$ is the function score computed with the coverage in place of the pair score.

\subsubsection{Match Propagation}
After each scoring round, we select confident source--binary pairs as seeds. For a pair $(s,b)$ to qualify, $b$ must score at least $\tau=0.5$ and lead all other candidates by at least $\delta=0.05$, while the paths of $s$ must contain at least eight operations in total.
This score already includes the caller and callee terms, with a smaller caller weight (\appautoref{app:context}).
For each seed, \sys rewrites every binary call to $b$ and every source call to $s$ into one shared call that names $s$, so that these calls carry information again.
Calls that keep their names are not rewritten.
The weights are recomputed on the rewritten paths, and all source functions are scored again.
This process runs for two rounds, and the caller and callee terms use the function scores $S$ computed after the second rewriting.

\subsubsection{Caller and Callee Terms}
The source and binary call graphs come from the calls on the extracted paths.
The caller term raises the score of every candidate called by the top candidate of a source caller of $s$ by the caller weight $\mu_c=0.15$ times the score of this top candidate.
The callee term raises the score of every candidate that calls the top candidate of a source callee of $s$ by the callee weight $\mu_e=0.05$.
Both terms use only callers and callees whose top candidate scores at least $\sigma=0.3$, and the sum of the function score and the two terms is the matching score $S'(s,b)$.
\appautoref{app:context} gives the formulas and further details.
\sys ranks candidates by $S'$ and breaks ties with the rules of \appautoref{app:ranking}.

\subsection{Detecting 1-Day Vulnerabilities}\label{sec:approach-rq4}

Detection must separate the vulnerable and the patched function, which share most of their behavior~\cite{zhang2018precise}.
The evidence therefore lies on the paths through the changed lines.
\sys compares these paths as full entry-to-exit paths rather than only the changed operations, because a few operations carry little information on their own after compilation.

\subsubsection{Function Localization}\label{sec:approach-localization}
For the vulnerable and for the patched source function, \sys selects the candidate whose name equals the source function name, ignoring case.
If no such candidate exists, \sys selects the candidate with the highest matching score.
If the selected candidate has a function score of at most 0, \sys reports no candidate for this source function.
If \sys reports a candidate for both source functions, the located function is the one with the higher function score (\appautoref{app:oneday}).

\subsubsection{Patch Presence Test}\label{sec:approach-patch}
The changed lines are the lines that differ between the vulnerable and the patched function.
In both functions, \sys selects the extracted paths through the changed lines.
Because the extracted paths may miss these lines, \sys also selects, for each reachable CPG node on these lines, the longest entry-to-exit path through this node.
The differing paths $D_v$ are the selected paths of the vulnerable function whose normalized sequences differ from all selected and extracted paths of the patched function, and the differing paths $D_p$ of the patched function are defined in the same way.
The extracted paths of the located function may also miss the compiled code of the change.
\sys therefore repeatedly adds the longest entry-to-exit path through an operation that no path covers yet, until every operation that can reach an exit with each loop body entered at most once (\autoref{sec:approach-paths}) lies on a path.

Each path of $D_v$ and $D_p$ then selects its best binary path in the located function.
This selection is a match if its pair score reaches a threshold $\theta$.
When several paths select the same binary path, only the highest-scoring path keeps the match.
Keeping only the highest-scoring path lets the changed operations decide between paths that share most of their operations.
\sys scores each of the two functions by the mean of two values, the number of its matches divided by $|D_v|+|D_p|$ and the mean pair score of these matches.
A function without matches scores 0.
The binary is classified as vulnerable if the vulnerable function scores at least as high as the patched function.

In \autoref{fig:motivation}, the vulnerable function has two differing paths, one through the branch with a dictionary, which has six ownership tests, and one through the branch without a dictionary.
The patched function has one differing path, which keeps one of the six ownership tests (\autoref{fig:approach-patch}).
On the vulnerable GCC \cc{-O2} target, a vulnerable path keeps the binary path that the patched path selects, and on the patched target, the patched path keeps it.
The classification is correct on all eight stripped target binaries of this vulnerability.

\subsection{Implementation}\label{sec:implementation}
We implement \sys in Python, and it uses no learned model.
\sys decompiles binaries with Ghidra~10.3~\cite{ghidra} and uses Ghidra~12.1.4 to recover the functions that version 10.3 misses.
Joern~4.0.623~\cite{joern} builds the code property graphs of both sides.
\sys uses the parameters in \autoref{tab:params}.
Because a path may enter each loop body only once (\autoref{sec:approach-paths}), path extraction tracks each CPG node together with the loops already entered, and it is limited to 250,000 such pairs and 120 seconds per function.
For 1-day detection, we set $\theta$ to 0.26, which maximizes Youden's $J$ statistic~\cite{youden1950index} on stripped targets (\appautoref{app:oneday}).
This calibration uses the ground-truth address of the function in each target, but not whether the target is vulnerable or patched.
All experiments ran on AMD EPYC nodes with 128 cores and 1~TB of RAM.

\section{Evaluation}

\label{sec:evaluation}

We answer the following research questions.
\squishlist
\item \textbf{RQ1 (\sbridge Replication):} 
How do query exclusions and tie handling affect the reported accuracy of \sbridge?
\item \textbf{RQ2 (Accuracy):} How accurately does \sys match source to binary functions under compiler, optimization, stripping, architecture, and inlining variations?
\item \textbf{RQ3 (Ablation Study):} How much does each design of \sys contribute to function matching accuracy?
\item \textbf{RQ4 (1-day Detection):} How effectively does \sys detect 1-day vulnerabilities?
\item \textbf{RQ5 (Scalability):} How does \sys scale with binary size and the number of candidate functions?
\squishend

\subsection{Experimental Setup}\label{sec:eval-setup}

\PP{Datasets}
\label{sec:impl-dataset}
We use datasets from \sbridge~\cite{yang2026sbridge} for function matching and 1-day vulnerability detection, respectively.
The function matching dataset consists of 122 programs from GNU Coreutils 9.1~\cite{coreutils} and Inetutils 2.4~\cite{inetutils}.
Each program is compiled for four architectures (x86, x64, ARM32, and ARM64) under eight configurations (GCC 9.4.0 or Clang 13.0, \cc{-O0} or \cc{-O2}, and stripped or unstripped), yielding 3,904 binaries in total.
\textit{The vulnerability detection dataset} covers 26 1-day CVEs across four projects: tcpdump~\cite{tcpdump}, OpenSSL~\cite{openssl}, libxml2~\cite{libxml2}, and FFmpeg~\cite{ffmpeg}.
We use vulnerable and patched releases of each project: tcpdump (4.9.0/4.9.3), OpenSSL (3.0.0/3.0.9), libxml2 (2.10.0/2.10.3), and FFmpeg (4.4/4.4.4).
Both releases are compiled for x64 under eight configurations (GCC 11.4.0 or Clang 14.0.6, \cc{-O0} or \cc{-O2}, and stripped or unstripped).
For each CVE, we evaluate 16 targets (eight vulnerable and eight patched) to determine whether each is correctly classified.

\PP{Ground truth}
We use the ground truth released with \sbridge, derived from symbol names and call graphs of unstripped binaries.
For function matching, source functions are mapped to binary functions with the same symbol names.
Functions absent from \cc{-O2} builds are considered inlined and mapped to their nearest surviving callers in the \cc{-O0} call graph, assuming no inlining at \cc{-O0}.
If an inlined function retains a separate copy with its own symbol, only that copy is annotated.
Inlined functions with no surviving callers are excluded due to the lack of ground truth.
We thus evaluate 64,016 function matching queries with ground truth, 32,008 each for stripped and unstripped binaries.
All function matching results are reported on these queries unless stated otherwise. For 1-day detection, the ground truth provides function addresses for 416 targets (208 stripped and 208 unstripped).

\PP{Baselines}
We compare \sys with two types of methods: 
source-to-binary function matching and source-level clone detection.
For source-to-binary matching, we compare with \sbridge \cite{yang2026sbridge}.
We examine its released implementation~\cite{sbridgeartifact} and 
identify issues in the matching procedure and evaluation that affect the reported accuracy.
To ensure a fair comparison, we replicate \sbridge (\autoref{sec:eval-rq1}) 
and evaluate both methods on the same queries with the same metrics.
For source-level clone detection, 
we evaluate VUDDY~\cite{kim2017vuddy} to assess whether source-level methods can be applied to decompiled code.
We apply VUDDY to source and decompiled functions.
VUDDY \emph{matches no query}.
It normalizes and hashes each function body, reporting a match only when two hashes are identical.
Decompiled code differs from the source in its statements and control structures, so no pair of hashes agrees.

\PP{Metrics}
A common metric for this task is Recall@$k$, which calculates the percentage of queries with a correct function among the top $k$ candidates.
This metric is ambiguous when multiple candidates have the same score, and a scoring method that cannot separate similar candidates can produce many such ties.
We therefore use a stricter metric, \textbf{\boldmath Recall$_s$@$k$}, which does not count ties as correct.
Recall$_s$@1 requires a correct function to have the unique top score.
Recall$_s$@5 requires the correct function and all candidates tied with it to fit in the first five positions.
We apply this metric to all evaluated methods.
For 1-day detection, we report precision, recall, and F1.
Following \sbridge, a target is considered located only if the selected function's address matches the ground truth; otherwise, it is classified as patched.
A target classified as vulnerable counts as a true positive if it is vulnerable and a false positive if it is patched.

\PP{Statistical analysis}
We assess the statistical significance of differences using 95\% confidence intervals computed from 3,000 paired bootstrap resamples.
Because the same source function can appear in multiple binaries, queries involving that function are not independent.
We resample by source function within each program suite for function matching, 
and by CVE for 1-day detection.
To account for multiple comparisons, 
we apply the Bonferroni correction~\cite{dunn1961multiple}, 
computing confidence intervals at the $1-0.05/m$ confidence level for each family of $m$ comparisons.
We consider a difference statistically significant if its confidence interval excludes zero.
All differences between percentages are absolute.

\subsection{RQ1: Replication of \sbridge}\label{sec:eval-rq1}

We built \sbridge from the released source and ran its function matching stage on the released features and ground truth for all 3,904 binaries.
We scored the output with the authors' scripts.
\sbridge reports Recall@1 rather than Recall$_s$@$k$.
Our run gives a Recall@1 of 75.27\% overall and 71.30\% on stripped binaries.\footnote{An implementation bug makes \sbridge return rankings for 90 queries that are not in the ground truth, and it mistakenly counts 82 of them as correct. We exclude these 90 queries in the rest of this section, which lowers 71.30\% to 71.23\%.}
We show the full comparison in \autoref{tab:rq1-table5} of \appautoref{app:replication}.
The results differ only slightly from Table~5 of the \sbridge paper, which reports 75.13\% and 71.19\%, respectively.
The Docker image of the \sbridge artifact also contains the authors' own run.
When we process the files in the order of their dataset archive, our output is identical to that run.
We are therefore confident that we ran the tool correctly.

We describe the two main issues of \sbridge below, and show in \appautoref{app:replication} the others.
Here we show their effects on the 32,008 stripped queries.

\PP{Issue 1: Unanswered queries are excluded}
The released ground truth contains 32,008 stripped queries, but \sbridge returns a ranking for only 25,772 of them, which we call the \emph{answered} queries.
\sbridge leaves out the other 6,236 queries (19.5\%).
These queries come from two kinds of source functions that \sbridge skips: those whose released control block features are empty (6,172 queries, from 387 functions) and those whose name begins with an underscore (the other 64).
Most of the 387 functions have empty features because of a limitation in the source front end of \sbridge.
It parses each function in isolation and fails on common declarations, such as those with a C23 attribute or with certain pointer return types.
For example, \sbridge extracts no features from \cc{find\_int}, a 32-line function of the Coreutils program \cc{test}.
Because \sbridge returns no results for these queries, they should be counted as misses rather than being excluded.
This would drop \sbridge's Recall@1 on stripped binaries from 71.23\% to 57.35\% (\autoref{tab:rq1-grid}).

\begin{figure}[t]
\centering
\includegraphics[width=0.65\linewidth]{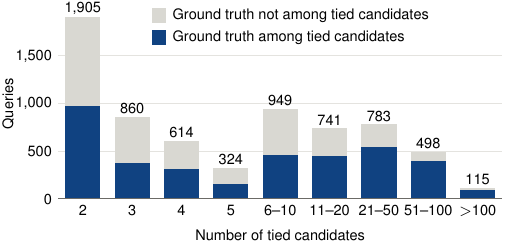}
\caption{Number of candidates that tie for the top score of \sbridge, on the 6,789 stripped queries with a tied positive top score. 
\sbridge counts the queries in the blue bars as correct.
}
\label{fig:rq1-ties}
\end{figure}

\PP{Issue 2: Tied candidates count as correct}
The \sbridge paper defines a correct match as the function ranked first, but several candidates often share the top score.
\sbridge counts a query as correct if one of these tied candidates is correct.
Of the 25,772 answered queries, 7,779 (30.18\%) have multiple top-ranked candidates.
In 990 of them, every candidate scores 0.00.
In 2,137 of them, more than ten candidates tie for a positive top score.
\autoref{fig:rq1-ties} shows the number of tied candidates for the 6,789 queries with a positive top score.
With ties counted, 18,358 of the 25,772 answered queries are correct, and 3,734 of them count as correct only because of ties.
The query of \autoref{sec:intro} with 185 tied candidates searches for \cc{nomoreargs} of the Coreutils program \cc{expr} in a stripped ARM64 \cc{-O0} binary built by Clang, and all 185 score 1.00.
Because an analyst cannot tell which tied candidate is correct, ties should not count as correct.
Under Recall$_s$@1, the result on stripped binaries drops further from 57.35\% to 45.69\%.

Ties also inflate Recall@5, because \sbridge applies the same rule to it.
If more than five candidates tie for the top score, the query counts as correct for Recall@5 whenever one of them is correct, although five positions cannot hold all tied candidates.

\begin{table}[t]
\caption{Replication results of \sbridge on stripped binaries. Each cell gives the number of correct queries and the recall. 
All: all queries with ground truth, where unanswered queries are misses.
}
\label{tab:rq1-grid}
\centering
\footnotesize
{%
\begin{tabular}{@{}lrr@{}}
\toprule
Query set & Recall@1 & Recall$_s$@1 \\
\midrule
\sbridge answered (25,772) & 18,358 (71.23\%) & 14,624 (56.74\%) \\
All (32,008) & 18,358 (57.35\%) & 14,624 (45.69\%) \\
\bottomrule
\end{tabular}%
}
\end{table}

\summarybox{
Our replication reveals several issues in \sbridge.
Fixing the two main ones, excluding unanswered queries and counting ties as correct, drops its accuracy on stripped binaries.
We fix all these issues in the rest of the evaluation.
We reported the issues to the authors of \sbridge, who acknowledged them.
}

\subsection{RQ2: Function Matching}\label{sec:eval-rq2}

We compare \sys with \sbridge on the queries that \sbridge answers and on all queries with ground truth.

\begin{table}[t]
\caption{Function matching on stripped and unstripped binaries (\%). The better result of each comparison is in bold. Differences are given with 95\% confidence intervals. 
 VUDDY~\cite{kim2017vuddy}, which matches no query, is not shown.
 }
\label{tab:rq2-overall}
\centering
\small
\maxsizebox{\textwidth}{!}
{%
\begin{tabular}{@{}llrrlrrl@{}}
\toprule
 & & \multicolumn{3}{c}{Stripped} & \multicolumn{3}{c}{Unstripped} \\
\cmidrule(lr){3-5}\cmidrule(l){6-8}
Query set & Metric & \sbridge & \sys & \sys $-$ \sbridge & \sbridge & \sys & \sys $-$ \sbridge \\
\midrule
\multirow{2}{*}{\sbridge answered (25,772)} & Recall$_s$@1 & 56.74 & \textbf{79.90} & +23.15 [+20.11, +26.26] & 65.16 & \textbf{80.90} & +15.75 [+13.23, +18.29] \\
 & Recall$_s$@5 & 72.25 & \textbf{88.46} & +16.20 [+13.77, +18.78] & 80.73 & \textbf{89.60} & +8.88 [+6.99, +10.92] \\
\midrule
\multirow{2}{*}{All (32,008)} & Recall$_s$@1 & 45.69 & \textbf{77.87} & +32.18 [+28.56, +35.71] & 52.46 & \textbf{79.14} & +26.68 [+23.23, +30.07] \\
 & Recall$_s$@5 & 58.18 & \textbf{87.31} & +29.13 [+25.51, +32.71] & 65.00 & \textbf{88.81} & +23.81 [+20.39, +27.29] \\
\bottomrule
\end{tabular}
}
\end{table}

\PP{Overall accuracy}
We show the overall results in \autoref{tab:rq2-overall}.
On all 32,008 stripped queries with ground truth, \sys reaches a Recall$_s$@1 of 77.87\%, against 45.69\% for \sbridge, and reduces the misses by 59\% (7,083 against 17,384).
Under Recall$_s$@5, \sys reaches 87.31\%, against 58.18\% for \sbridge.
The lead has two sources.
First, \sys is more accurate on the 25,772 queries that \sbridge answers, with a Recall$_s$@1 of 79.90\% against 56.74\%.
Even when we measure \sbridge with Recall@1, which counts its ties as correct, and \sys with Recall$_s$@1, \sys leads with 79.90\% against 71.23\%.
Second, \sbridge returns no result for 6,236 queries, and \sys is correct on 4,334 of them (69.50\%).
\sys also covers almost all correct matches of \sbridge, as it is correct on 13,869 of the 14,624 queries on which \sbridge is correct (94.84\%).

\begin{table}[t]
\caption{Function matching by group on all 32,008 queries with ground truth. The better result of each comparison is in bold. 
The intervals are at the 95\% level. 
Due to space limits, differences and intervals are given for Recall$_s$@1 only.
}
\label{tab:rq2-config}
\centering
\small
\maxsizebox{\textwidth}{!}
{%
\begin{tabular}{@{}lrrrrrrrrrrr@{}}
\toprule
 & & \multicolumn{5}{c}{Stripped} & \multicolumn{5}{c}{Unstripped} \\
\cmidrule(lr){3-7}\cmidrule(l){8-12}
 & & \multicolumn{3}{c}{Recall$_s$@1 (\%)} & \multicolumn{2}{c}{Recall$_s$@5 (\%)} & \multicolumn{3}{c}{Recall$_s$@1 (\%)} & \multicolumn{2}{c}{Recall$_s$@5 (\%)} \\
\cmidrule(lr){3-5}\cmidrule(lr){6-7}\cmidrule(lr){8-10}\cmidrule(l){11-12}
Group & Queries & \sbridge & \sys & Difference & \sbridge & \sys & \sbridge & \sys & Difference & \sbridge & \sys \\
\midrule
ARM32 & 7,954 & 40.11 & \textbf{71.35} & +31.24 [+27.40, +34.80] & 54.38 & \textbf{81.68} & 49.03 & \textbf{74.11} & +25.08 [+21.74, +28.30] & 63.16 & \textbf{85.44} \\
ARM64 & 8,016 & 47.16 & \textbf{80.09} & +32.93 [+28.92, +36.79] & 60.93 & \textbf{90.12} & 52.33 & \textbf{80.10} & +27.77 [+24.31, +31.27] & 65.74 & \textbf{90.00} \\
x86 & 8,016 & 45.21 & \textbf{77.66} & +32.45 [+29.10, +35.70] & 55.58 & \textbf{86.12} & 53.77 & \textbf{79.94} & +26.17 [+22.59, +29.68] & 64.92 & \textbf{88.64} \\
x64 & 8,022 & 50.24 & \textbf{82.34} & +32.10 [+28.26, +36.01] & 61.79 & \textbf{91.26} & 54.69 & \textbf{82.37} & +27.69 [+24.27, +31.29] & 66.16 & \textbf{91.15} \\
\midrule
GCC & 16,040 & 45.57 & \textbf{78.10} & +32.54 [+29.01, +35.90] & 57.99 & \textbf{87.71} & 52.61 & \textbf{78.90} & +26.29 [+22.98, +29.55] & 65.21 & \textbf{88.74} \\
Clang & 15,968 & 45.81 & \textbf{77.64} & +31.83 [+27.91, +35.66] & 58.37 & \textbf{86.90} & 52.32 & \textbf{79.39} & +27.07 [+23.49, +30.64] & 64.79 & \textbf{88.88} \\
\midrule
\cc{-O0} & 16,053 & 55.65 & \textbf{89.25} & +33.61 [+29.26, +37.93] & 63.78 & \textbf{95.20} & 62.54 & \textbf{89.48} & +26.94 [+23.25, +30.70] & 70.75 & \textbf{95.36} \\
\cc{-O2} & 15,955 & 35.67 & \textbf{66.42} & +30.75 [+27.35, +34.02] & 52.54 & \textbf{79.37} & 42.33 & \textbf{68.74} & +26.42 [+23.03, +29.83] & 59.21 & \textbf{82.23} \\
\midrule
Coreutils & 21,041 & 45.90 & \textbf{74.79} & +28.90 [+24.13, +33.47] & 60.02 & \textbf{85.27} & 53.66 & \textbf{75.26} & +21.61 [+17.63, +25.65] & 67.61 & \textbf{85.85} \\
Inetutils & 10,967 & 45.29 & \textbf{83.78} & +38.49 [+34.93, +42.09] & 54.65 & \textbf{91.21} & 50.17 & \textbf{86.59} & +36.42 [+32.70, +40.30] & 60.00 & \textbf{94.49} \\
\midrule
Total & 32,008 & 45.69 & \textbf{77.87} & +32.18 [+28.56, +35.71] & 58.18 & \textbf{87.31} & 52.46 & \textbf{79.14} & +26.68 [+23.23, +30.07] & 65.00 & \textbf{88.81} \\
\bottomrule
\end{tabular}
}
\end{table}

\PP{Compiler, architecture, and optimization}
On stripped binaries, \sys leads in every group of \autoref{tab:rq2-config}, by 28.90\% (Coreutils) to 38.49\% (Inetutils) in Recall$_s$@1.
After the Bonferroni correction for these eleven comparisons, including the total, the lead of \sys remains significant in every group.
The compiler has little effect on \sys, with a difference of 0.46\% between GCC and Clang.
Architecture has a larger effect, with Recall$_s$@1 ranging from 71.35\% on ARM32 to 82.34\% on x64.
Optimization has the largest effect on both tools.
At \cc{-O2}, \sys nearly doubles the Recall$_s$@1 of \sbridge, with 66.42\% against 35.67\%.

\PP{Inlining}
Inlining is the hardest case for both tools.
A query is inlined if the ground truth marks its source function as inlined in the binary that the query searches.
We show the results on the 4,899 inlined queries that \sbridge answers in \autoref{tab:rq2-inline} of \appautoref{app:replication}.
Under Recall$_s$@1, \sys reaches 49.23\%, against 28.03\% for \sbridge.
Under Recall@1, which counts ties as correct for both tools, \sbridge leads with 55.38\% against 49.85\%.
Both differences are significant.
However, in 1,341 of the 2,713 queries on which \sbridge is correct under Recall@1, the correct function ties for the top score with other candidates, and the median tied group has nine candidates.

\PP{Unstripped binaries}
On unstripped binaries, a symbol lookup alone resolves 26,145 of the 32,008 queries, so these binaries are a weak test of matching.
To measure matching rather than name lookup, \sys builds the paths of each unstripped binary from its stripped copy.
Stripping removes only the symbols, so the two copies contain the same machine code.
If \sys decompiled the unstripped binary instead, Ghidra would write the names of internal functions and global variables into the code, and calls would match the source by name.
\sys uses the symbol table only for function entries, which extend the candidate lists (\appautoref{app:unstripped}).
\sys reaches a Recall$_s$@1 of 79.14\%, against 52.46\% for \sbridge, whose features keep the names of internal callees.

\PP{Failure cases}
Of the 7,083 stripped queries that \sys misses, 3,020 are near misses, with the correct function among the first five candidates.
Most misses are at \cc{-O2} (5,358).
We assign each miss to the first of four causes that applies.
In 1,135 misses, the correct function is not among the candidates, or its path extraction did not complete.
In 2,652, the source function is inlined and keeps no separate copy.
In 1,467, the source function is small, with at most ten operations on its paths.
In 124, the top candidate is the counterpart of a near-duplicate of the source function. Two source functions are near-duplicates if the operations on their paths have a multiset Jaccard similarity of at least 0.9.
The other 1,705 misses have none of these causes.

\summarybox{
With a Recall$_s$@1 of 77.87\% against 45.69\% on stripped binaries, \sys leads \sbridge overall and in every group.
It is more accurate on the queries that \sbridge answers, even when \sbridge is measured with Recall@1, and it is correct on 4,334 queries that \sbridge cannot answer.
On inlined queries, the hardest case, \sys still leads under Recall$_s$@1.
}

\subsection{RQ3: Ablation Study}\label{sec:eval-rq3}

We ablate four key designs of \sys, namely behavioral paths, containment, ordered alignment, and calling context refinement.
To measure the effect of paths, we start from two baselines without paths.
They represent a function as a set of CPG nodes and differ only in the score.
Let $O_s$ and $O_b$ be the sets of normalized operations of all CPG nodes of a source function $s$ and a binary function $b$, with the operation normalization of \sys.
\squishlist
\item \emph{CPG nodes + Jaccard} uses neither paths nor containment. It scores $b$ by the Jaccard index $|O_s \cap O_b| / |O_s \cup O_b|$, which penalizes operations missing on either side.
\item \emph{CPG nodes + containment} adds containment but no paths. It scores $b$ by $|O_s \cap O_b| / |O_s|$, the share of source operations found in $b$, so extra code in $b$ does not lower the score.
\squishend

Starting from CPG nodes + containment, we then add the designs of \sys one at a time.
\squishlist
\item \emph{Path containment} adds paths, together with operation weights, operation multiplicities, partial credit for wrapper functions, and the length factor. It scores the containment of source paths in binary paths, regardless of the order of operations (\autoref{sec:approach-scoring}).
\item \emph{+ Ordered alignment} tests whether the order of operations matters. It aligns the operations of each path pair in order, with the gap cost.
\item \emph{+ Inline-host rule}, \emph{+ Match propagation}, and \emph{+ Caller and callee terms} test whether calling context helps. They add the three steps of calling context refinement in turn (\autoref{sec:approach-propagation}). 
\squishend
The Inlined and Other columns split the \cc{-O2} queries that \sbridge answers.

\begin{table}[t]
\caption{Ablation study of \sys on the 
stripped queries.
}
\label{tab:rq3-ablation}
\centering
\small
\setlength{\tabcolsep}{3pt}
\maxsizebox{\columnwidth}{!}
{
\begin{tabular}{@{}lrrrrrr@{}}
\toprule
 & \multicolumn{5}{c}{Recall$_s$@1} & Recall$_s$@5 \\
\cmidrule(lr){2-6}\cmidrule(l){7-7}
Variant & All & \cc{-O0} & \cc{-O2} & Inlined & Other & All \\
\midrule
CPG nodes + Jaccard & 58.80 & 77.26 & 40.23 & 6.90 & 62.34 & 70.05 \\
CPG nodes + containment & 56.66 & 63.91 & 49.37 & 47.76 & 52.62 & 75.21 \\
Path containment & 63.59 & 74.89 & 52.22 & 35.50 & 65.42 & 78.01 \\
+ Ordered alignment & 70.54 & 83.62 & 57.39 & 34.48 & 75.27 & 81.73 \\
+ Inline-host rule & 72.97 & 83.37 & 62.50 & 48.91 & 74.44 & 84.48 \\
+ Match propagation & 75.08 & 85.96 & 64.13 & \textbf{49.97} & 76.50 & 85.60 \\
+ Caller and callee terms & \textbf{77.87} & \textbf{89.25} & \textbf{66.42} & 49.23 & \textbf{80.55} & \textbf{87.31} \\
\bottomrule
\end{tabular}
}
\end{table}

\PP{Node-set baselines}
We show the results in \autoref{tab:rq3-ablation}.
A set of operations is not enough to match functions.
CPG nodes + Jaccard reaches a Recall$_s$@1 of 58.80\%, and only 6.90\% on inlined queries.
A symmetric score penalizes the code around an inlined function, so this baseline rarely finds its host.
CPG nodes + containment raises Recall$_s$@1 on inlined queries to 47.76\%, which shows that containment is necessary for them.
However, its overall result falls to 56.66\%, and its result at \cc{-O0} falls from 77.26\% to 63.91\%.
Even CPG nodes + Jaccard exceeds \sbridge (45.69\%), because it uses the operation normalization of \sys, answers every query, and produces few ties.
The designs of \sys add another 19.07\% on top of this baseline.

\PP{Paths and containment}
Paths raise accuracy once containment is in place.
Path containment raises Recall$_s$@1 to 63.59\%, 6.93\% above CPG nodes + containment, and its result at \cc{-O0} to 74.89\%.
This step lowers Recall$_s$@1 on inlined queries to 35.50\%, which the inline-host rule later recovers.
In \sys itself, replacing containment with a symmetric score, which divides by the mean weight of both paths, costs 11.73\% (not shown in \autoref{tab:rq3-ablation}).
Weighting paths by the behavior they carry also helps, because equal path weights lower Recall$_s$@1 by 0.43\%.

\PP{Ordered alignment}
Ordered alignment adds 6.95\% and raises Recall$_s$@1 to 70.54\%.
This gain does not reach inlined queries.
Ordered alignment needs a sequence, which a set of nodes lacks.
In \sys itself, replacing it with the order-free score of Path containment costs 5.13\%.

\PP{Calling context refinement}
Calling context refinement adds 7.33\% overall and 14.75\% on inlined queries.
The inline-host rule raises Recall$_s$@1 on inlined queries to 48.91\%.
Match propagation and the caller and callee terms raise Recall$_s$@1 to 75.08\% and 77.87\%.
After the Bonferroni correction, every gain in the All column remains significant.
The same holds for unstripped binaries, where these steps raise Recall$_s$@1 from 59.59\% to 79.14\% (\autoref{tab:rq3-ablation-unstripped} in \appautoref{app:unstripped}).

\summarybox{
Each design of \sys is necessary, and together they raise Recall$_s$@1 from 58.80\% to 77.87\%.
Containment and the inline-host rule are key for inlined queries, and ordered alignment adds 6.95\%.
}

\subsection{RQ4: 1-Day Vulnerability Detection}\label{sec:eval-rq4}

We follow the evaluation of \sbridge.
Its inputs have two defects, which we fix for both tools (E4 and E5 in \autoref{tab:rq1-issues} in \appautoref{app:replication}).
The fixes improve the results of \sbridge.

\begin{table}[t]
\caption{1-day detection. Each row has 104 vulnerable and 104 patched targets (208 each in All), so TP+FN and FP+TN are 104 (208). Correct is TP+TN. 
}
\label{tab:rq4-main}
\centering
\small
\setlength{\tabcolsep}{2pt}
{%
\begin{tabular}{@{}ll|rrrrr|rrr@{}}
\toprule
Targets & \multicolumn{1}{l}{Tool} & TP & FP & TN & FN & \multicolumn{1}{r}{Correct} & Prec. & Rec. & F1 \\
\midrule
\multirow{2}{*}{Stripped} & \sbridge & 44 & 26 & 78 & 60 & 122 & 0.63 & 0.42 & 0.51 \\
 & \sys & \textbf{66} & \textbf{18} & \textbf{86} & \textbf{38} & \textbf{152} & \textbf{0.79} & \textbf{0.63} & \textbf{0.70} \\
\midrule
\multirow{2}{*}{Unstripped} & \sbridge & 65 & 42 & 62 & 39 & 127 & 0.61 & 0.63 & 0.62 \\
 & \sys & \textbf{83} & \textbf{29} & \textbf{75} & \textbf{21} & \textbf{158} & \textbf{0.74} & \textbf{0.80} & \textbf{0.77} \\
\midrule
\multirow{2}{*}{All} & \sbridge & 109 & 68 & 140 & 99 & 249 & 0.62 & 0.52 & 0.57 \\
 & \sys & \textbf{149} & \textbf{47} & \textbf{161} & \textbf{59} & \textbf{310} & \textbf{0.76} & \textbf{0.72} & \textbf{0.74} \\
\bottomrule
\end{tabular}%
}
\end{table}

\PP{Overall results}
We show the results in \autoref{tab:rq4-main}.
\sys substantially outperforms \sbridge on stripped targets.
\sys detects 50\% more vulnerable targets (66 against 44) with 31\% fewer false positives (18 against 26).
As a result, it improves both precision (0.79 against 0.63) and recall (0.63 against 0.42), and its F1 score is 0.70, against 0.51 for \sbridge.
\sys locates the function in 158 of the 208 targets, against 143 for \sbridge.
It correctly classifies 152 targets, against 122 for \sbridge, a lead of 14.4\% with a 95\% confidence interval from 3.8\% to 25.0\%.

On unstripped targets, \sys classifies 158 targets correctly, against 127 for \sbridge, and reaches an F1 score of 0.77 against 0.62.
On all 416 targets, the scope of Table~7 of the \sbridge paper, \sys reaches an F1 score of 0.74 against 0.57.
After the Bonferroni correction, the lead of \sys in correct classifications remains significant on stripped targets, on unstripped targets, and on all targets.

\begin{table}[t]
\caption{1-day detection by project on the stripped targets, and F1 on stripped and unstripped targets.
}
\label{tab:rq4-oss}
\centering
\small
\setlength{\tabcolsep}{3pt}
{%
\begin{tabular}{@{}lr|rr|rr@{}}
\toprule
 & & \multicolumn{2}{c}{Correct} & \multicolumn{2}{c}{F1 (all)} \\
\cmidrule(lr){3-4}\cmidrule(l){5-6}
Project (\#CVE) & Targets & \sbridge & \multicolumn{1}{r}{\sys} & \sbridge & \sys \\
\midrule
tcpdump (10) & 80 & 51 & \textbf{59} & 0.60 & \textbf{0.75} \\
OpenSSL (8) & 64 & 32 & \textbf{52} & 0.54 & \textbf{0.81} \\
libxml2 (2) & 16 & 11 & \textbf{13} & 0.67 & \textbf{0.84} \\
FFmpeg (6) & 48 & 28 & 28 & 0.52 & \textbf{0.54} \\
\midrule
Total (26) & 208 & 122 & \textbf{152} & 0.57 & \textbf{0.74} \\
\bottomrule
\end{tabular}%
}
\end{table}

\PP{Results by project}
We show the results by project in \autoref{tab:rq4-oss}.
\sys correctly classifies more stripped targets in tcpdump, OpenSSL, and libxml2.
In OpenSSL, it classifies 52 targets correctly, against 32 for \sbridge.
In FFmpeg, both tools classify 28 targets correctly, but \sbridge detects 12 of the 24 vulnerable targets and \sys only 4.
\sbridge uses a string filter that its paper does not describe, which compares a source function only with binary functions that share one of its strings (E6 in \autoref{tab:rq1-issues}).
On 16 FFmpeg targets, this filter reduces the candidate list to at most three functions.
\sys instead ranks every candidate and does not rely on such a filter.

\summarybox{
Under the evaluation of \sbridge and with corrected inputs, \sys reaches an F1 score of 0.70 on stripped targets, against 0.51 for \sbridge, with higher precision and recall, and 0.74 against 0.57 on all targets.
\sys classifies more stripped targets correctly in three of four projects and ties on FFmpeg.
}

\subsection{RQ5: Scalability}\label{sec:eval-rq5}

We measure the time to match one source function against one binary.
With a single source function, \sys has no calling context to refine and uses the function scores alone.
The time excludes decompilation, CPG construction, and path extraction, which \sys runs once before any query, as \sbridge excludes its preprocessing.
To keep repeated runs of \sbridge short, we compare the two tools on 12 programs from the function matching dataset, with the 1,008 queries that \sbridge answers in their 192 stripped binaries, which have 33 to 261 candidates.
Both tools run three times on the same AMD EPYC 7702 node with one thread, and each value is the median over the three runs.
As in the released implementation of \sbridge, the time of the first query of a binary includes loading its candidates.

We also time \sys alone in a single run on all 32,008 stripped queries and on the 32 stripped binaries of the 1-day dataset, eight per project.
Here we report the time per query without loading, and the load time separately.
Loading reads the paths of all candidates of a binary into memory.
For the 1-day dataset, a query matches one source function of the benchmark against one binary of its project.

\begin{table}[t]
\caption{Time of \sys on stripped binaries with the function scores alone. Queries gives the total and Candidates the range. All other values are medians. Load is the one-time cost to load the candidates of a binary.}
\label{tab:rq5-scale}
\centering
\footnotesize
\setlength{\tabcolsep}{2pt}
\maxsizebox{\columnwidth}{!}
{%
\begin{tabular}{@{}lrrrrrr@{}}
\toprule
 & Size & & & & Query & Load \\
Binaries & (MiB) & Candidates & Queries & Ops. & (s) & (s) \\
\midrule
Function matching & 0.06 & 33--558 & 32,008 & 38 & 0.05 & 1.3 \\
tcpdump & 1.42 & 1,023--1,636 & 2,432 & 139 & 1.01 & 64.4 \\
libxml2 & 1.61 & 2,275--2,843 & 32 & 1,296 & 4.75 & 63.0 \\
OpenSSL & 4.63 & 2,342--13,287 & 192 & 84 & 1.32 & 168.9 \\
FFmpeg & 20.75 & 17,146--18,951 & 96 & 715 & 62.96 & 1,186.5 \\
\bottomrule
\end{tabular}%
}
\end{table}

\PP{Speed and scale}
\sys is 5.7 times faster than \sbridge on average.
On the 1,008 queries, \sys takes 0.38~s per query on average and 0.05~s at the median, whereas \sbridge takes 2.16~s on average and 1.22~s at the median.
The share of these queries that take less than 10~s, which the \sbridge paper also reports, is 100\% for \sys and 98.6\% for \sbridge.
With calling context refinement, \sys takes 0.41~s per query over all 1,280 queries with ground truth in these binaries, including loading.
These 1,280 queries include the 272 that \sbridge does not answer.
\autoref{tab:rq5-scale} shows the results on all stripped binaries, where Ops. is the number of normalized operations on the paths of a source function.
Across programs, the time of \sys without loading grows with the number of candidates and with the size of the source function.
On binaries with up to 13,287 candidates, every query takes at most 18~s without loading.
FFmpeg, whose binaries and source functions are both large, is the slowest, with a median of about one minute.
Loading one FFmpeg binary takes about 20~min at the median.

\summarybox{
\sys is 5.7 times faster than \sbridge. Without loading, queries take at most 18~s on binaries with up to 13,287 candidates, and on FFmpeg the median query takes about one minute.
}

\section{Discussion and Limitations}\label{sec:discussion}

\PP{Failure cases}
At \cc{-O2}, the counterpart of an inlined function without a separate copy is a host function.
The additional operations of the host lower the length factor, and the inline-host rule raises the host only when it is the top candidate of a caller of the function.
Inlining the callee into the source paths of its callers would give \sys source paths of the same shape as the host.
A small function has few operations, which many unrelated candidates also perform, so its paths carry little evidence.
Its callers and callees could carry more weight in its score.
The near-duplicates \cc{add\_fs\_type} and \cc{add\_excluded\_fs\_type} of \cc{df} differ only in the global variable that they update, which a stripped binary does not name.
Propagating matches through global variables, as match propagation does through calls, could separate them.
Finally, function recovery can miss a counterpart, and path extraction can fail on large functions or on control flow that it does not model.
Combining decompilers and adding a fallback extraction for such functions could close part of this gap.

\PP{Patch presence test}
Containment cannot show that a path is absent.
One fix in tcpdump adds a source path that shares most of its operations with a path of the vulnerable function, so two vulnerable stripped \cc{-O0} binaries are classified as patched.
Binary paths also carry no line information, so a match does not show that the compiled code of the change lies on the matched path.
PS3~\cite{zhan2024ps3} and REACT~\cite{zhan2024react} test the code of the change itself, so running either on the function that \sys locates could address both limitations.
Each needs the function to be located first (\autoref{sec:related}).

\PP{Scope of the evaluation}
Our results are limited to the compilers, optimization levels, architectures, and projects of the two datasets of \sbridge (\autoref{sec:impl-dataset}).
All projects are written in C, and no binary is obfuscated.
Control flow flattening would hide the order of operations that ordered alignment relies on, and inserted bogus code would lower the length factor as inlined code does.
RQ4 also keeps the ground-truth check of \sbridge for localization, so it does not measure detection without this check.

\section{Related Work}\label{sec:related}

\PP{Source-to-binary matching}
Source-to-binary matching has mostly been used for software composition analysis (\autoref{sec:background-similarity}).
OSSPolice~\cite{duan2017oss} and B2SFinder~\cite{yuan2019b2sfinder} match features such as strings, constants, and exported functions against source repositories.
BinaryAI~\cite{jiang2024binaryai} matches binary and source functions by the similarity of embeddings from a learned model, refines the matches with link-time locality, and reports the reused components.
\sbridge~\cite{yang2026sbridge} decompiles the binary and matches the control blocks (\autoref{sec:problem-prior}) of a source function with those of each binary function.
XLoc~\cite{shi2026xloc} uses a large language model to rewrite a source function into decompiler-like code and then accepts or rejects each candidate binary function.
However, existing techniques either only identify the reused component, compare features that compilation and inlining change, or rely on a large language model, and none of them matches the behavior that compilation preserves.

\PP{1-day vulnerability detection}
1-day vulnerability detectors find reused vulnerable code in source code or test whether a binary contains a patch (\autoref{sec:background-similarity}).
VUDDY~\cite{kim2017vuddy} matches hashes of normalized functions, MOVERY~\cite{woo2022movery} and MVP~\cite{xiao2020mvp} match signatures derived from vulnerabilities and their patches, and V1SCAN~\cite{woo2023v1scan} combines version-based and code-based detection.
For binaries, patch presence tests such as FIBER~\cite{zhang2018precise} and PS3~\cite{zhan2024ps3} compile reference binaries from the source code and test whether a binary contains signatures of the patch.
FIBER derives these signatures with binary symbolic execution, and PS3 derives them with symbolic emulation so that they stay stable across compilers and optimization levels.
REACT~\cite{zhan2024react} instead compiles the vulnerable and the patched source code to an intermediate representation and compares both with the representation lifted from the binary.
However, existing techniques either match the syntax of the source code that they scan, which decompiled code does not reproduce, or need the symbol or the address of the vulnerable function, which a stripped binary does not reveal for an internal function and which binary-to-binary matching recovers only with a reference binary.

\section{Conclusion}\label{sec:conclusion}
We presented \sys, which locates source functions in stripped binaries through behavioral containment.
\sys decompiles the binary, represents each function by behavioral paths, aligns source paths with binary paths in order, and refines the scores with calling context.
Over all stripped queries, with ties counted as misses, it ranks the correct function first for 77.87\%, against 45.69\% for \sbridge.
It still leads when \sbridge is measured as in its original evaluation, with 79.90\% against 71.23\%.
On stripped binaries, its F1 score for 1-day vulnerability detection is 0.70, against 0.51 for \sbridge.
Inlined and small functions cause most of the remaining misses, which source paths with inlined callees and more weight on calling context could reduce.

\section*{Acknowledgments}
This work was supported in part by the NTU Start-Up Grant and the Ministry of Education, Singapore, under its Academic Research Fund Tier 1 (RS8/26). 
Yifei Yang is supported in part by the Jiangxi Provincial Natural Science Foundation (Grant Nos.~20252BAC240082, 20262BAC200476, 20262BAC240423, 20262BEJ730020) and the Jiangxi Provincial Key Laboratory of Data Security Technology (Grant No.~20242BCC32026), and gratefully acknowledges additional financial support from the China Scholarship Council (CSC).
Penghui Li is the corresponding author.
\bibliographystyle{unsrt}
\bibliography{references,conf}

\clearpage

\appendix

\section{Method Details}\label{app:method}
\label{app:supplementary}
\subsection{Path Extraction}\label{app:path}
Before parsing, the binary side needs a syntax normalization that adds type definitions for the built-in type names of Ghidra and rewrites a small number of constructs of decompiled code that are not valid C, such as scoped names.
Non-returning library functions come from a fixed list, such as \cc{exit}, \cc{abort}, and \cc{error} with a nonzero integer literal as status.
A function defined in the program is also non-returning if its control flow has no cycle and ends only in calls to non-returning functions.
A call to \cc{error} with an unknown status is both an exit and an ordinary call.

\subsection{\texorpdfstring{Operation Weights}{Operation Weights}}\label{app:weights}
If $M$ is the number of paths of all binaries under analysis and $\mathit{df}(t)$ is the number of these paths that contain operation $t$, the weight of $t$ is
\begin{equation}\label{eq:idf}
  w(t) = \log\frac{M+1}{\mathit{df}(t)+1} + 1\,.
\end{equation}
Weights are computed per program, separately over its stripped and unstripped binaries.
A source operation that occurs in no path of the queried binary receives weight~0 for that binary, because no candidate can match it.

\subsection{Calling Context Refinement}\label{app:context}
Let $\hat{g}$ denote the top candidate of a source function $g$ under $S$, with ties resolved by a fixed order of the recovered functions, and let $a\rightarrow b$ denote that binary function $a$ calls binary function $b$.
Let $\mathcal{R}(s)$ be the set of source callers $g$ of $s$ with $S(g,\hat{g})\geq\sigma$, where $\sigma=0.3$.
The caller term increases the scores of the candidates called by the top candidate of each caller in $\mathcal{R}(s)$:
\begin{equation}\label{eq:ccs}
  S_1(s,b) = S(s,b) + \mu_c \sum_{g\in\mathcal{R}(s)} S(g,\hat{g})\,[\hat{g}\rightarrow b]\,,
\end{equation}
where $\mu_c=0.15$ is the caller weight and $[\cdot]$ is 1 if the condition holds and 0 otherwise.

The callee term raises the score of every candidate that calls the top candidate of a source callee of $s$.
In the callee term, $\hat{f}$ denotes the top candidate of $f$ under $S_1$, and $\mathcal{E}(s)$ is the set of source callees $f$ of $s$ with $S_1(f,\hat{f})\geq\sigma$.
With the callee weight $\mu_e=0.05$, the matching score is
\begin{equation}\label{eq:ces}
  S'(s,b) = S_1(s,b) + \mu_e \sum_{f\in\mathcal{E}(s)} [b\rightarrow\hat{f}]\,.
\end{equation}

The inline-host rule runs once per scoring round, determines all top candidates before raising any score, and ignores direct recursive calls.
In seed selection, each candidate has at most one seed, the source function with the highest score in that round, and seeds from earlier rounds are kept.
The caller and callee terms in this selection use $\hat{g}$ and $\hat{f}$ from the function scores of that round.
Because an inlined callee has no call from its host, the caller term raises the functions that the host calls instead of the host.

\subsection{Final Ranking}\label{app:ranking}
Ties in the final ranking are broken in favor of the candidate $b$ with the most source callers of $s$ that are seeds for a candidate that calls $b$.
Remaining ties are broken by the smaller difference in the total number of operations on the paths of $s$ and $b$ and finally by a fixed pseudorandom order.
Recall$_s$@$k$ does not use this pseudorandom order, so candidates that the two rules above leave tied count as tied.

\subsection{1-Day Detection}\label{app:oneday}
If an unstripped binary has no function named after the source function, \sys first selects the highest-scoring function named after a source caller, because an inlined function can leave no symbol.
When \sys reports a candidate for both source functions, it keeps the one with the higher function score and resolves a tie in favor of the vulnerable function.
When it reports none, the binary is not located.
If a change has no line on one side of the patch, that side uses the lines around the change.
To calibrate $\theta$, the positives are the pair scores of paths common to the vulnerable and the patched function on the correctly located function, and the negatives are the pair scores of these paths on 50 other functions chosen at random.
We use the same value on unstripped targets.

\begin{figure}[t]
\centering
\includegraphics[width=0.67\linewidth]{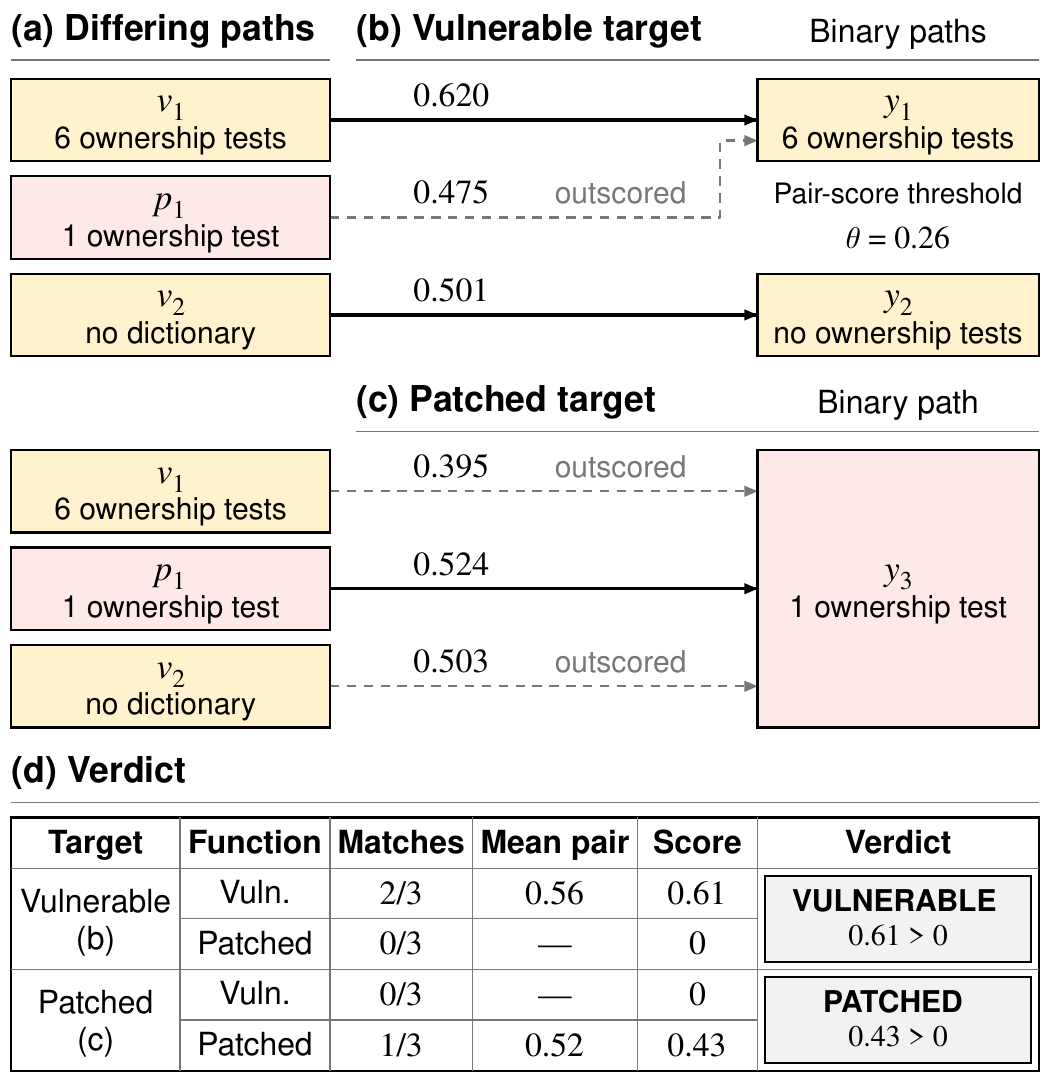}
\caption{Patch presence test of \cc{xmlFreeEntity} on the vulnerable and the patched GCC \cc{-O2} target. (a) The differing paths are the paths through the changed lines that the other function lacks, namely $v_1$ and $v_2$ of the vulnerable function and $p_1$ of the patched function. An ownership test checks whether the dictionary of \autoref{fig:motivation} owns a string, and $v_2$ takes the branch without a dictionary. In (b) and (c), numbers on arrows are pair scores, and a solid arrow marks the path that keeps the match. (b) On the vulnerable target, the patched path passes the threshold but loses the shared binary path to a higher-scoring vulnerable path. (c) On the patched target, all three paths select the same binary path, and the patched path keeps the match. (d) The score of a function is the mean of its match fraction and its mean pair score.}
\label{fig:approach-patch}
\end{figure}

\begin{table}[ht!]
\centering
\caption{Parameters of \sys.}
\label{tab:params}
\small
\begin{tabularx}{\columnwidth}{@{}lXr@{}}
\toprule
Symbol & Description & Value \\
\midrule
$\lambda$ & Cost of an unmatched binary operation & 0.1 \\
$\kappa$ & Partial credit for wrapper functions & 0.4 \\
$\beta$ & Exponent of the length factor & 0.15 \\
$\rho$ & Inline-host factor & 0.9 \\
$\tau$ & Minimum score of a seed & 0.5 \\
$\delta$ & Minimum score margin of a seed & 0.05 \\
$\sigma$ & Minimum score of a caller or callee match & 0.3 \\
$\mu_c$ & Caller weight in the final ranking & 0.15 \\
 & Caller weight before seed selection & 0.1 \\
$\mu_e$ & Callee weight & 0.05 \\
 & Minimum number of operations of a seed & 8 \\
 & Propagation rounds & 2 \\
$\theta$ & Patch presence test threshold & 0.26 \\
\bottomrule
\end{tabularx}
\end{table}

\section{Additional Details of the \sbridge Replication}\label{app:replication}
\autoref{tab:rq1-issues} includes the two issues of \autoref{sec:eval-rq1} (E1 and E2).

\PP{Function matching}
Of the 90 queries without ground truth that \sbridge ranks (\autoref{sec:eval-rq1}), 82 count as correct because an empty ground truth matches every candidate name.
If the first listed candidate is taken as the answer, the result depends on the order in which \sbridge enumerates the candidates.
Across eight orders, it ranges from 47.65\% to 50.45\% of the stripped queries with ground truth.
Counting ties as correct also rewards loose matching.
Before we exclude the 90 queries, lowering the block similarity threshold of \sbridge from 0.7 to 0.2 raises its stripped Recall@1 from 71.30\% to 73.99\%.
However, the share of answered queries with a single top candidate falls from 69.76\% to 59.33\%.

\PP{1-day detection}
FN in the \sbridge paper appears to include targets that are not located, because TP+FN is 275, although only 208 targets are vulnerable.
Our run on the released inputs leaves 69 patched targets not located, close to the 67 extra FNs.
We therefore recount Table~7 of the \sbridge paper.
With the corrected inputs of \autoref{sec:eval-rq4} (E4 and E5), the F1 score of \sbridge on the 416 targets rises from 0.4629 to 0.5662, close to the 0.5757 that the counts of the \sbridge paper give.
Because the cells of \autoref{tab:rq1-table7} still differ, we cannot determine which inputs the authors used.

\begin{table}[ht!]
\caption{Replication of Table~5 of the \sbridge paper with the authors' scoring script. Stripped and unstripped binaries are combined unless the row states otherwise. R@$k$ is Recall@$k$ in percent, and MRR is the mean reciprocal rank.}
\label{tab:rq1-table5}
\centering
\small
\begin{tabular}{@{}lcccccc@{}}
\toprule
 & \multicolumn{3}{c}{\sbridge paper} & \multicolumn{3}{c}{Our run} \\
\cmidrule(lr){2-4}\cmidrule(l){5-7}
Group & R@1 & R@5 & MRR & R@1 & R@5 & MRR \\
\midrule
ARM32 & 70.82 & 77.87 & .7718 & 71.13 & 77.90 & .7741 \\
ARM64 & 77.51 & 83.25 & .8275 & 77.68 & 83.25 & .8283 \\
x86 & 73.94 & 79.21 & .7869 & 73.86 & 79.20 & .7862 \\
x64 & 78.27 & 83.58 & .8299 & 78.43 & 83.62 & .8308 \\
\midrule
GCC & 75.37 & 80.97 & .8051 & 75.44 & 80.93 & .8054 \\
Clang & 74.90 & 80.99 & .8029 & 75.11 & 81.06 & .8043 \\
\midrule
\cc{-O0} & 84.96 & 87.54 & .8802 & 85.14 & 87.64 & .8811 \\
\cc{-O2} & 65.31 & 74.41 & .7279 & 65.41 & 74.35 & .7286 \\
\midrule
Unstripped & 79.09 & 84.70 & .8375 & 79.25 & 84.62 & .8385 \\
Stripped & 71.19 & 77.25 & .7706 & 71.30 & 77.37 & .7712 \\
\midrule
Total & 75.13 & 80.98 & .8040 & 75.27 & 80.99 & .8048 \\
\bottomrule
\end{tabular}
\end{table}

\begin{table}[ht!]
\caption{Recount of Table~7 of the \sbridge paper on the 416 targets of its 1-day dataset. As in \autoref{sec:eval-setup}, a target that is not located is classified as patched, so TP+FN and FP+TN are 208. We recompute the row of the \sbridge paper from its TP and FP. The paper itself reports an F1 score of 0.4936.}
\label{tab:rq1-table7}
\centering
\small
\setlength{\tabcolsep}{5pt}
\maxsizebox{\columnwidth}{!}{%
\begin{tabular}{@{}lrrrrccc@{}}
\toprule
Inputs & TP & FP & TN & FN & P & R & F1 \\
\midrule
\sbridge paper & 116 & 79 & 129 & 92 & .5949 & .5577 & .5757 \\
Released inputs & 78 & 51 & 157 & 130 & .6047 & .3750 & .4629 \\
+ rebased addresses & 90 & 65 & 143 & 118 & .5806 & .4327 & .4959 \\
+ missing functions & 109 & 68 & 140 & 99 & .6158 & .5240 & .5662 \\
\bottomrule
\end{tabular}%
}
\end{table}

\begin{table}[ht!]
\caption{The six most important issues in the implementation and evaluation of \sbridge. E1 to E3 concern the function matching dataset. E4 to E6 concern the 1-day dataset, and we report their effects on its 208 stripped targets.}
\label{tab:rq1-issues}
\centering
\small
\begin{tabularx}{\textwidth}{@{}lXX@{}}
\toprule
ID & Issue & Measured effect \\
\midrule
E1 & Queries with ground truth that \sbridge cannot answer are excluded from Recall@1. & \sbridge excludes 6,236 of the 32,008 stripped queries with ground truth. Counting them as misses lowers Recall@1 from 71.23\% to 57.35\%. \\
E2 & Recall@1 counts every candidate tied for the top score at two decimal places as correct, although the \sbridge paper defines a correct match as the function ranked first. & Not counting ties lowers 57.35\% to 45.69\% (Recall$_s$@1) on the 32,008 stripped queries with ground truth. Recall@5 follows the same rule. \\
E3 & The reported overall Recall@1 averages stripped and unstripped binaries, and a symbol lookup alone resolves most unstripped queries. & The \sbridge paper reports 75.13\% overall and 71.19\% on stripped binaries. The lookup resolves 26,145 of the 32,008 unstripped queries. \\
\midrule
E4 & The ground truth keeps the addresses of the ELF files, while Ghidra rebases position-independent executables and shared objects. & The ground-truth check of 1-day detection would reject the correct function on 58 targets. Rebasing raises located targets from 87 to 130 and the F1 score from 0.33 to 0.42. \\
E5 & The candidate lists of seven builds are incomplete. & Adding the missing functions raises located targets from 130 to 143 and the F1 score from 0.42 to 0.51. \\
E6 & \sbridge compares a source function that has a string block only with binary functions that share one of its strings, when such functions exist. Its paper does not describe this step. & The filter is active on 104 targets, and on 13 of them the ground truth is never compared. \\
\bottomrule
\end{tabularx}
\end{table}

\begin{table}[ht!]
\caption{Function matching on the inlined queries of stripped \cc{-O2} binaries that \sbridge answers, in the layout of Table~6 of the \sbridge paper (\%). Share: inlined queries among the answered \cc{-O2} queries. 
}
\label{tab:rq2-inline}
\centering
\small
\maxsizebox{\columnwidth}{!}{%
\begin{tabular}{@{}lrrrrrr@{}}
\toprule
 & & & \multicolumn{2}{c}{Recall@1} & \multicolumn{2}{c}{Recall$_s$@1} \\
\cmidrule(lr){4-5}\cmidrule(l){6-7}
Group & Queries & Share & \sbridge & \sys & \sbridge & \sys \\
\midrule
Coreutils Clang & 2,436 & 54.12\% & 56.16 & 53.49 & 27.55 & 52.91 \\
Coreutils GCC & 2,059 & 45.65\% & 52.70 & 44.10 & 27.10 & 43.42 \\
Inetutils Clang & 222 & 11.33\% & 65.77 & 64.86 & 34.68 & 64.86 \\
Inetutils GCC & 182 & 9.29\% & 62.64 & 47.80 & 36.81 & 46.70 \\
\midrule
Total & 4,899 & 37.89\% & 55.38 & 49.85 & 28.03 & 49.23 \\
\bottomrule
\end{tabular}%
}
\end{table}

\section{Function Matching on Unstripped Binaries}\label{app:unstripped}
\begin{table}[ht!]
\caption{Ablation study of \sys on the 32,008 unstripped queries (Recall$_s$@1 and Recall$_s$@5, \%). The columns are those of \autoref{tab:rq3-ablation}.}
\label{tab:rq3-ablation-unstripped}
\centering
\small
\setlength{\tabcolsep}{3pt}
\maxsizebox{\columnwidth}{!}
{
\begin{tabular}{@{}lrrrrrr@{}}
\toprule
 & \multicolumn{5}{c}{Recall$_s$@1} & Recall$_s$@5 \\
\cmidrule(lr){2-6}\cmidrule(l){7-7}
Variant & All & \cc{-O0} & \cc{-O2} & Inlined & Other & All \\
\midrule
CPG nodes + Jaccard & 59.59 & 77.16 & 41.91 & 6.90 & 64.37 & 71.10 \\
CPG nodes + containment & 57.53 & 63.94 & 51.09 & 48.44 & 54.50 & 76.26 \\
Path containment & 64.48 & 75.07 & 53.82 & 35.64 & 67.39 & 79.15 \\
+ Ordered alignment & 71.67 & 83.72 & 59.54 & 34.70 & 77.90 & 83.07 \\
+ Inline-host rule & 74.00 & 83.48 & 64.46 & 48.87 & 76.96 & 85.76 \\
+ Match propagation & 76.14 & 86.09 & 66.14 & \textbf{49.77} & 79.12 & 87.01 \\
+ Caller and callee terms & \textbf{79.14} & \textbf{89.48} & \textbf{68.74} & 49.17 & \textbf{83.67} & \textbf{88.81} \\
\bottomrule
\end{tabular}
}
\end{table}
The released candidate lists lack the ground truth of 235 queries.
We therefore extend the lists of \sys with the functions that the symbol table defines in executable sections and the candidates of the stripped copy of the same build.
The symbol table thus gives \sys function entries, though not names.
\sbridge has no features for the added candidates and keeps the released lists.
The lead of \sys is 26.68\%, or 26.37\% without the 235 queries, and all eleven unstripped intervals of \autoref{tab:rq2-config} exclude zero at level $0.05/11$.
On the queries that \sbridge answers, the lead falls to 3.63\% when ties count as correct for both tools.
It falls to 1.72\% when ties count as correct for \sbridge only, and this lead is not significant.

\autoref{tab:rq3-ablation-unstripped} repeats the ablation of \autoref{sec:eval-rq3} on the unstripped queries.
Every row from Path containment raises Recall$_s$@1 in the All column, and every gain is significant after the Bonferroni correction.
As on stripped binaries, ordered alignment does not help inlined queries, and the inline-host rule raises Recall$_s$@1 on them from 34.70\% to 48.87\%.
In \sys itself, replacing containment with a symmetric score costs 11.87\%, replacing ordered alignment with the order-free score of Path containment costs 5.36\%, and equal path weights cost 0.52\%.

\end{document}